\documentclass[11pt,a4paper]{article}
\pdfoutput=1
\usepackage{jheppub}

\usepackage{subcaption}

\newcommand{\ba}{\begin{eqnarray}}
\newcommand{\ea}{\end{eqnarray}}

\newcommand{\barray}{\begin{array}}
\newcommand{\earray}{\end{array}}

\newcommand{\rd}{\mathrm{d}}
\newcommand{\re}{\mathrm{e}}

\def\del{\partial}

\title{Rotating kinky braneworlds}

\author{Florian Niedermann,}
\author{Paul M. Saffin}

\affiliation{School of Physics and Astronomy, University Park, University of Nottingham,\\ Nottingham NG7 2RD, United Kingdom}

\emailAdd{florian.niedermann@nottingham.ac.uk}
\emailAdd{paul.saffin@nottingham.ac.uk}

\abstract{Cylindrical braneworlds have been used in the literature as
  a convenient way to resolve co-dimension-two branes. They are
  prevented from collapsing by a massless worldvolume field with
  non-trivial winding, but here we discuss another way of preventing
  collapse, which is to rotate the brane. We use a simple microscopic
  field theory model of a domain wall with a condensate for which
  rotation is a necessity, not just a nice added extra. This is due to
  a splitting instability, whereby the effective potential trapping
  the condensate is not strong enough to hold it on the defect in the
  presence of winding without charge.

  We use analytic defect solutions in the field theory (kinky vortons)
  to construct a thin-wall braneworld model by including gravitational
  dynamics, and we allow for the rotation required by the microscopic
  theory. We then discuss the impact rotation has on the bulk and
    brane geometry, thereby providing an anchor for further cosmological
    investigations.  Our setup naturally leads to
  worldvolume fields living at slightly different radii, and we
  speculate on the consequences of this in regard to the fermion mass-hierarchy.}


\arxivnumber{1801.09983}

\begin{document}

\maketitle

\section{Introduction}
\label{sec:intro}
According to the braneworld idea, all Standard Model matter fields,
except for gravity, are confined to a brane that is localized inside a
higher dimensional bulk spacetime. This explains why our human
experience is limited to only three spatial dimensions, even though
additional large (or even infinite) extra dimensions might
exist. Since gravity has kinetic support in the bulk, the
presence of extra dimensions changes the way sources localized on the
brane gravitate. This modified gravitational sector is useful because
it provides a phenomenological window to infer the presence of extra
dimensions and at the same time it leads to a new perspective on
longstanding gravitational puzzles. Importantly, these aspects can be
studied by using an effective field theory approach to braneworlds
that remains agnostic about the model's high energy origin. In that context,
co-dimension-two braneworlds turned out to be particularly interesting
for a variety of reasons:
\begin{itemize}
\item \emph{The cosmological constant (CC) problem:} in 4D gravity the
  vacuum energy of Standard Model matter fields acts as a cosmological
  constant and thus destabilizes a Minkowski vacuum (or de Sitter
  vacuum with phenomenologically small curvature scale) unless some
  fine-tuning is imposed. This is the essence of the CC
  problem~\cite{Zeldovich:1967gd, Zeldovich:1968zz, Weinberg1989,
    Burgess:2013ara,Padilla:2015aaa}. Models with two infinite volume
  extra dimensions offer a built-in mechanism to hide the vacuum
  energy of Standard Model particles from a brane observer. Rather
  than producing a 4D de Sitter phase on the brane, vacuum energy
  deforms the bulk into a cone leaving the brane curvature
  unaffected. From the perspective of a brane observer, the vacuum
  energy is therefore decoupled. Unfortunately, it has not yet been
  possible to exploit this mechanism in a phenomenologically viable
  way. Specifically, the \textit{brane induced gravity model}
  (BIG)~\cite{Dvali:2000xg,Kaloper:2007ap}, which is based on the DGP
  mechanism~\cite{Dvali:2000hr} to restore a 4D gravity regime, was
  found to suffer from ghost instablities in the relevant parameter
  regime~\cite{Dubovsky:2002jm,Hassan:2010ys,Niedermann:2014bqa,
    Eglseer:2015xla}. In fact, the authors
  in~\cite{Niedermann:2017cel} have argued recently that the failure
  can be understood from a model independent perspective. Their
  analysis uses a very general spectral decomposition of the
  gravitational propagator. They have also highlighted a small handful
  of loopholes to their obstructions, and it remains to be seen
  whether those can be realized within an extra dimensional
  context. At the end of this work, we will speculate on this
  possibility in light of our new results.
  
\item \emph{The hierarchy problem:} according to the ADD
  proposal~\cite{ArkaniHamed:1998rs,Antoniadis:1998ig} models with
  large (but finite) volume extra dimensions can offer a geometrical
  explanation for the weakness of gravity, which becomes a consequence
  of the large extra space volume. Models with two co-dimensions are
  particularly interesting as they are the most predictive: if the bulk
  gravitational scale is of order of $\sim \,10\, \mathrm{TeV}$ (or
  above), a ten mircron (or smaller) sized extra dimension is required
  in order to realize the observed coupling strength of the
  gravitational zero mode (unless there are substantial warping
  effects). This implies both signatures of quantum gravity in
  collider experiments as well as deviations from the Newtonian
  inverse square law in table top experiments. The
  \textit{supersymmetric large extra dimensions} (SLED)
  proposal~\cite{Aghababaie:2003wz} (see also~\cite{Gibbons:2003di})
  is a prominent example of such a 6D model and can be understood as
  the low energy version of a particular supergravity
  theory.\footnote{A later version of this
    model~\cite{Burgess:2011mt,Burgess:2011va} has also been claimed
    to address the CC problem. However, it was shown recently that it
    cannot prevent a parameter tuning~\cite{Niedermann:2015via,
      Niedermann:2015vbk}, questioning its prospects as a solution to
    the CC problem.}
  
\item \emph{Brane cosmology:} 6D models provide a minimal
  playground to study cosmological signatures of the braneworld
  paradigm. Note that this program has been realised in five
  dimensions in the case of the Randall-Sundrum
  model~\cite{Randall:1999ee,Randall:1999vf} (see
  also~\cite{Maartens:2010ar} for a review on brane cosmology) but
  little work has been done in higher dimensions due to a simple
  physical reason: in general, a brane with more than one co-dimension
  acts as an antenna of gravitational waves if the brane undergoes
  cosmological
  evolution~\cite{Niedermann:2014yka,Niedermann:2014bqa}.\footnote{Previous
    works typically neglect the brane back-reaction on the bulk
    geometry or work in an effective 4D picture which fails at early
    times when the Hubble length might drop below the size of the extra
    space~\cite{Maartens:2010ar}.} This in turn makes it difficult to
  describe the full time-dependent, coupled system of bulk-brane
  equations consistently. However, it is this type of new dynamical
  feature that might also lead to a rich phenomenology, both for
  finite and infinite volume models.
\end{itemize}

Despite their phenomenological prospects, co-dimension-two braneworlds
suffer from a technical difficulty which is absent in one
co-dimension: spacetime curvature diverges at the position of an
infinitely thin co-dimension-two brane. This can be understood as the
gravitational analogue of a charged string in electrodynamics for
which the Coulomb field diverges logarithmically. In the case of a
pure tension brane this is a rather mild conical singularity, which can
be modelled in terms of a two-dimensional delta-function. However,
this is no longer possible for cosmological brane matter with FLRW
symmetries~\cite{Vinet:2004bk}. This problem is typically dealt with
by blowing up the transverse brane directions. A particularly popular
choice consists in replacing the string-like brane by a hollow
cylinder (or ring from a purely extra dimensional
perspective).\footnote{An alternative possibility consists in smearing
  the brane fields over a disc. A corresponding microscopic model
  which uses a Nielsen-Olesen vortex~\cite{Nielsen:1973cs} as its
  blueprint was discussed recently and applied to the SLED proposal
  in~\cite{Burgess:2015nka,Burgess:2015gba}.} This is convenient
because then the brane becomes locally a co-dimension-one object,
making Israel's covariant matching techniques~\cite{Israel:1966rt}
applicable. This type of brane model was first introduced
in~\cite{Peloso:2006cq} to regularize a flux-stabilized, compact
(rugby ball shaped) 6D model but later also applied to models with
infinite volume extra dimensions~\cite{Kaloper:2007ap,
  Eglseer:2015xla}. In a more generic context, it was used
in~\cite{Burgess:2008yx} to derive (renormalized) matching conditions
of co-dimensions-two braneworlds. In order to prevent the brane's
compact direction from collapsing, a massless scalar field is added to
its worldvolume theory. The scalar winds around the brane, thereby
providing the angular pressure needed to stabilize the cylinder's
radial direction. Building up on these result, later work often used
an angular pressure component of the brane energy-momentum tensor to
effectively implement this stabilisation mechanism, rather than
resolving it in terms of a worldvolume scalar field (see for example
\cite{Niedermann:2014bqa, Niedermann:2014vaa, Niedermann:2015vbk}).

This work is guided by the question as to whether it is possible to
consistently embed the hollow cylinder construction in a (classical)
mircrophysical theory that resolves the brane at high energies. In
order to maintain a high level of generality, we study this setup
within a minimal extra dimensional framework, which should make it
possible to extent our construction to established finite (e.g.\ the
rugby ball
models~\cite{Peloso:2006cq,Aghababaie:2003wz,Niedermann:2015vbk}) and
infinite (e.g.\ the BIG model~\cite{Kaloper:2007ap}) volume braneworld
models, or to further study it in its minimal form in accordance with
the inflating ``cigar'' proposal
in~\cite{Niedermann:2014yka}. Specifically, we will use a domain wall
solution that is bent into a cylinder in order to model the brane
sector. A cylindrical collapse can then be avoided by localizing a
condensate inside the wall (or brane equivalently). These
configurations are known as \textit{kinky
  vortons}~\cite{Battye:2008zh,Battye2009a,Battye:2009nf} and were
studied in two spatial dimensions as a proxy for closed loops of
superconducting cosmic strings in three dimensions (so-called
\textit{vortons}~\cite{Davis:1988jq}). Due to the trapping of the
condensate the brane carries winding and charge, which both contribute
towards its stability. In fact, for kinky vortons it is
known~\cite{Battye:2008zh} that the charge is a vital ingredient to
avoid non-axial instabilities, whereby the potential trapping the
condensate is not strong enough to hold on to it.

The important realization of Sec.~\ref{sec:fieldTheoryVortons} is that
the condensate's N\" other charge leaves a low energy fingerprint in the
form of a non-vanishing angular momentum of the brane, and hence
cannot be characterized by an angular pressure (or winding)
alone. Correspondingly, the requirement of having a consistent
microphysical description in terms of a kinky vorton forces us to
include angular momentum in our low energy description. This is an
important observation for two reasons: Firstly, it introduces rotation
as a new dynamical feature of 6D braneworlds which to our knowledge
has not been considered before, and secondly, it raises the question
as to whether alternative ultraviolet (UV) embeddings might lead to
the same conclusion, making it hence imperative to include rotation
to the low energy description.

While answering this UV-sensitive question goes beyond the scope of
the present work, we devote Sec.~\ref{sec:thin_brane} to a first
discussion of the implications rotation has on the spacetime
geometry. This is done within a thin-wall approximation that makes
explicit contact with the established models based on a worldvolume
description of the brane. We are able to derive explicit solutions of
the coupled gravitational system, including the bulk Einstein and
brane matching equations. We then find two types of solutions; those
for which the bulk geometry asymptotes to a cone (sub-critical) and
those for which the bulk closes in a second axis
(super-critical). These solutions can serve as an anchor for more
detailed cosmological investigations of both finite and infinite
volume scenarios. In particular, in the non-rotating limit the
super-critical branch of our braneworld model has recently been shown
to feature a new mechanism to realize an inflationary phase on the
brane~\cite{Niedermann:2014yka} and thus holds greater
phenomenological promise.

We will conclude our work in Sec.~\ref{sec:conclusion} by discussing
different directions of future research, and possible implications for the fermion mass hieracrchy.

\section{Kinky vortons}
\label{sec:fieldTheoryVortons}
The idea behind a kinky vorton is to construct a domain wall (kink) in
two spatial dimensions, and then bend it into a circle. This, however,
would be unstable, as the tension in the kink would cause the radius
of the ring to reduce, and the kink would eventually collapse and
decay into radiation. To prevent this from happening we may put a
condensate on the kink, and arrange it such that the condensate
stabilizes the vorton at some radius, with the condensate providing
both N\"other charge and winding number to contribute towards
stability. We now discuss such a model, as described in
\cite{Battye:2008zh, Battye:2009nf}.

\subsection{Microphysical description}
\begin{figure}
  	\centering
	\begin{tabular}{lr}
	\hspace{-1.5cm}
	\includegraphics[width=0.4\textwidth]{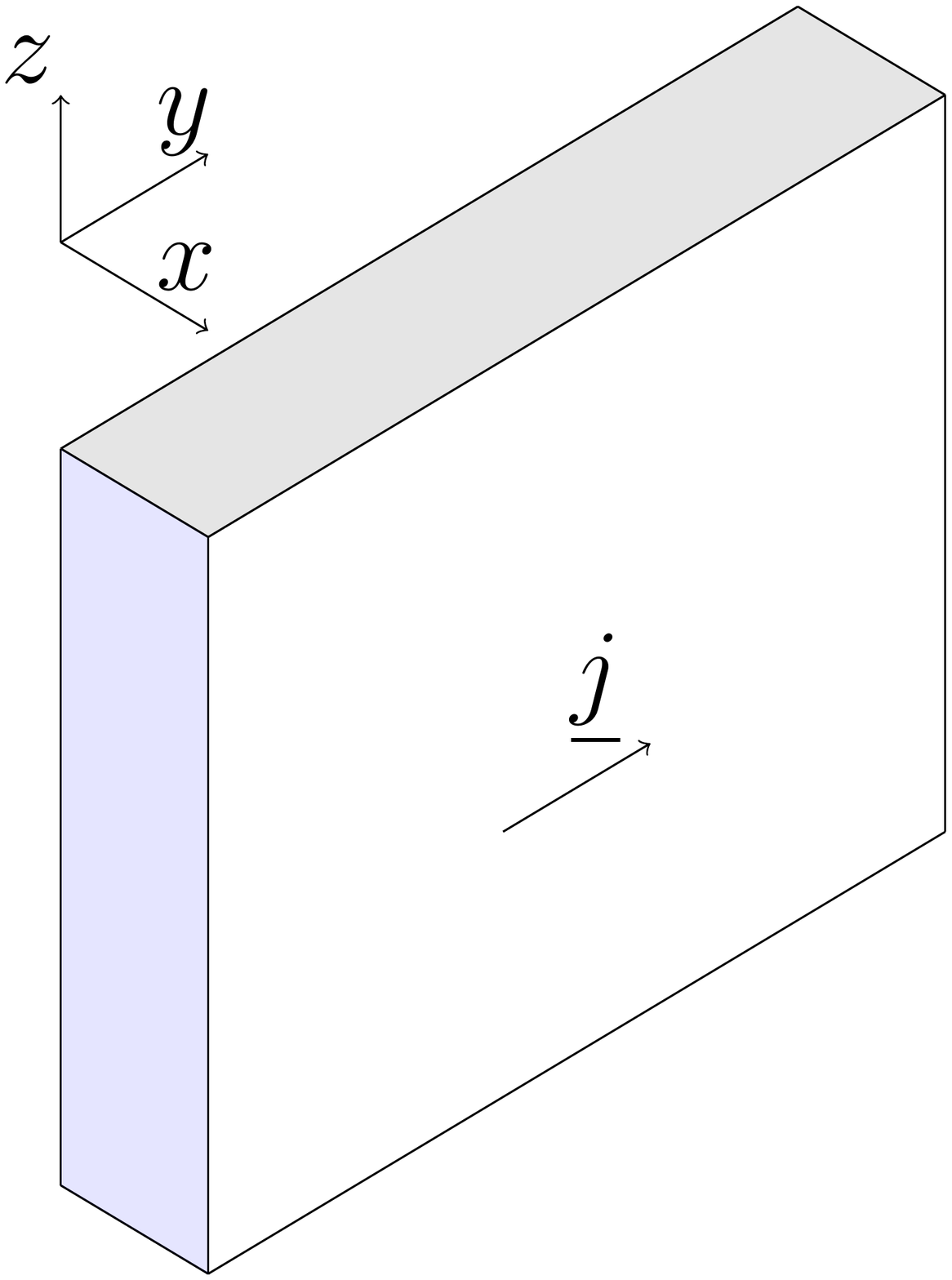} & \hspace{-.2cm}
	\includegraphics[width=0.4\textwidth]{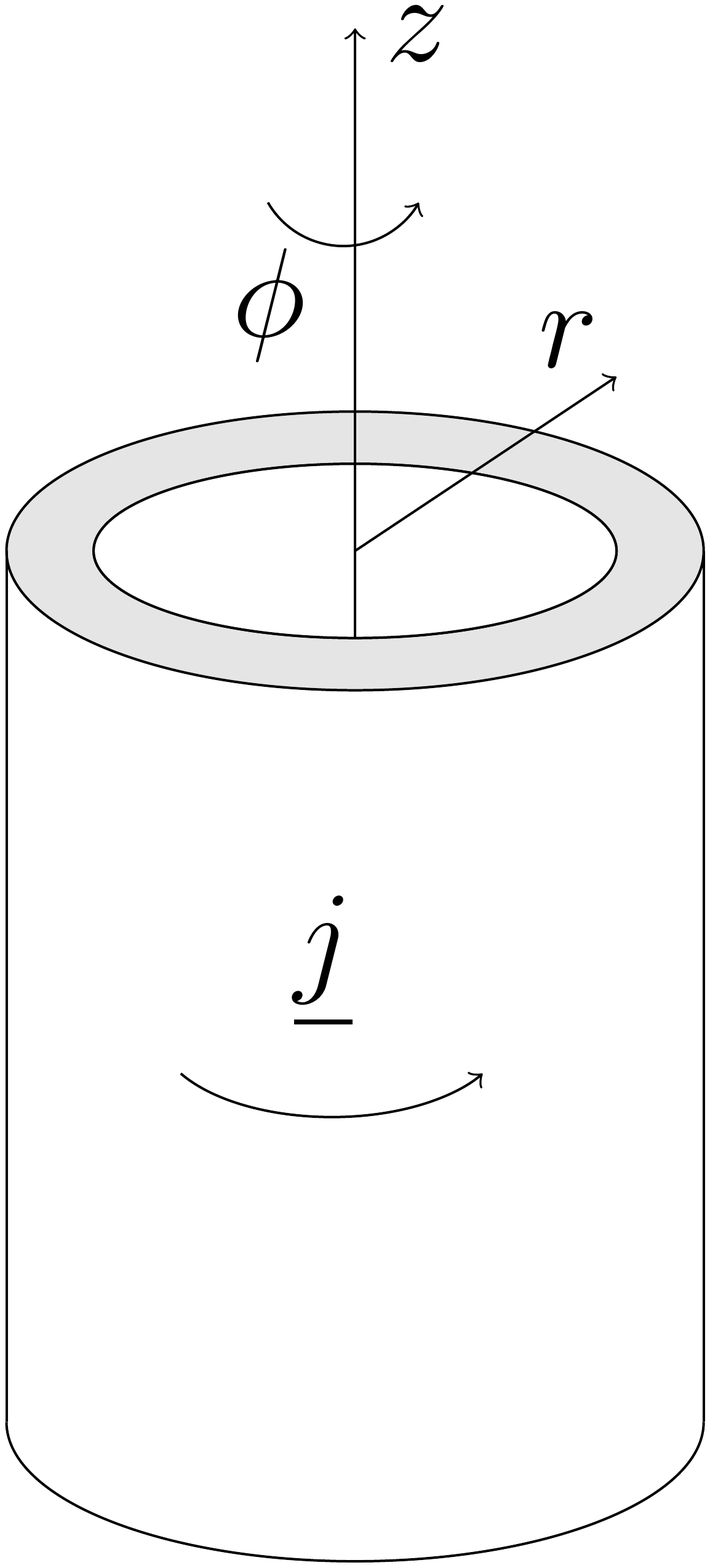}
	\end{tabular}
  	\caption{A sketch of the orientation of the domain wall and the vorton, each with current~$\underline{j}$.}\label{fig:sketch}
\end{figure}

The field responsible for the domain wall is taken to be a real scalar field
$\Phi$ with the standard symmetry-breaking potential, and we take the
condensate field to be a complex scalar $\Sigma$, again with the
standard symmetry-breaking form. The two fields have a bi-quadratic
coupling leaving us with \ba\nonumber {\cal
  L}&=&-\frac{1}{2}\, \del_\mu\Phi\del^\mu\Phi-\del_\mu\Sigma^\dagger\del^\mu\Sigma-\frac{1}{4} \, \lambda_\Phi \, \left(\Phi^2-\eta_\Phi^2\right)^2-\frac{1}{4} \, \lambda_\Sigma \left(|\Sigma|^2-\eta_\Sigma^2\right)^2 - \beta\,\Phi^2\,|\Sigma|^2
 + \frac{1}{4} \, \lambda_\Sigma \, \eta^4_\Sigma.\\\label{eq:fieldTheoryFullLagrangian}
\ea We shall take our model to live in $d$ spacetime dimensions, and
label the co-ordinates such that the fields are independent of $z$,
with $x$ being the transverse co-ordinate, and the current being in
the $y$-direction. Later, when we come to the circular vorton, then
the radial co-ordinate $r$ will play the role of $x$, and the
azimuthal direction $\phi$, along which the current flows, will play
the role of $y$, and we shall keep the notation $z$ for those
directions along which the fields do not depend (see
Fig.~\ref{fig:sketch}).

The constant term in the Lagrangian density is to ensure that $V(\Phi=\pm\eta_\Phi,\;\Sigma=0)=0$. The condensate is taken to be of the form
\ba\label{eq:straightCondensate}
\Sigma&=&|\Sigma|(x)\exp\{i\sigma(t,y)\}=|\Sigma|(x)\exp\{i[\omega t+ky]\},
\ea
and we define the quantity $\chi$, which is used to characterize vortons as electric ($\chi>0$), chiral ($\chi=0$) or magnetic ($\chi<0$),
\ba\label{eq:chiDefinition}
\chi&=&\omega^2-k^2.
\ea
The magnitude of the condensate field, $|\Sigma|$, is to vanish away from the kink, and take some non-zero value on the kink. In order to achieve this we note that the effective potential for the condensate in the core of the kink ($\Phi=0$) is given by
\ba\label{eq:sigmaEffectivePotential}
V(\Phi=0,\Sigma)&=&\frac{1}{4}\lambda_\Sigma\left(|\Sigma|^2-\eta^2_\Sigma-\frac{2\chi}{\lambda_\Sigma}\right)^2+C,\\
C&=&-\frac{1}{4}\lambda_\Sigma(\eta^2_\Sigma+2\chi/\lambda_\Sigma)^2+\frac{1}{4}\lambda_\Phi\eta^4_\Phi
\ea
and so for a non-zero $\Sigma$-condensate to form, we require
\ba
\eta_\Sigma^2+\frac{2\chi}{\lambda_\Sigma}>0,
\ea
and then the condensate has value $|\Sigma|^2=\eta_\Sigma^2+\frac{2\chi}{\lambda_\Sigma}$, which minimizes $V$.

In order to ensure the spontaneous breaking of the $\mathbb Z_2$ symmetry $\Phi\to-\Phi$, and hence the formation of a kink, we impose that the global vacua are $(\Phi=\pm\eta_\Phi,\;\Sigma=0)$, and so require the condensate on the kink to have an energy penalty compared to the global vacuum, \mbox{$V(\Phi=0,\;|\Sigma|^2=\eta_\Sigma^2+\frac{2\chi}{\lambda_\Sigma})>V(\Phi=\pm\eta_\Phi,\;\Sigma=0)$} . This leads to
\ba
\label{eq:lambdaPhiInequality}
\eta^4_\Phi\frac{\lambda_\Phi}{\lambda_\Sigma}>\left(\eta_\Sigma^2+2\chi/\lambda_\Sigma\right)^2.
\ea
Finally, we wish to ensure that no condensate forms when $\Phi=\pm\eta_\Phi$, and so we note that a positive quadratic term in
\ba
V(\Phi=\pm\eta_\Phi,\;\Sigma)&=&\left(\beta\eta^2_\Phi-\frac{1}{2}\lambda_\Sigma\eta^2_\Sigma-\chi\right)|\Sigma|^2+\frac{1}{4}\lambda_\Sigma|\Sigma|^4+constant
\ea
would prevent such a breaking of the U(1) symmetry by the presence of a condensate, which necessitates 
\ba
\label{eq:betaInequality}
\beta>\frac{\lambda_\Sigma}{2}\frac{\eta^2_\Sigma+2\chi/\lambda_\Sigma}{\eta^2_\Phi}.
\ea
At this point we make a parameter choice that will allow us to find analytic solutions, by imposing that (\ref{eq:lambdaPhiInequality}) and (\ref{eq:betaInequality}) are equivalent. This leads to
\ba
\label{eq:betaRelation}
\beta=\frac{1}{2}\sqrt{\lambda_\Phi\lambda_\Sigma}.
\ea
The analytic solution in question is
\begin{subequations}
\ba
\label{eq:analyticProfiles}
\Phi&=&\eta_\Phi\tanh(ax),\\
\label{eq:analyticProfiles_sigma}
|\Sigma|&=&\frac{b}{\cosh(ax)},\\
a^2&=&\beta\eta_\Phi^2-(\chi+\lambda_\Sigma\eta_\Sigma^2/2),\\
b^2&=&\frac{2}{\lambda_\Sigma}(2(\chi+\lambda_\Sigma\eta_\Sigma^2/2)-\beta\eta_\Phi^2),
\ea
\end{subequations}       
which further requires
\ba
\lambda_\Sigma&=&4\beta,
\ea
and then (\ref{eq:betaRelation}) gives $\lambda_\Phi=\beta$. Having found the analytic solution for a kink with a condensate, we need to ensure that the solution parameters $a$ and $b$ are real, leading to
\ba
2\beta\eta_\Sigma^2+\chi<\beta\eta_\Phi^2<2(2\beta\eta_\Sigma^2+\chi).
\ea
At this point we make another choice of parameters, which ensures a symmetric range in possible values of $\chi$
\ba\label{eq:choice}
\eta_\Phi^2&=&\frac{8}{3}\eta_\Sigma^2,
\ea
leading to
\ba
-1<4\tilde\chi<1,
\ea
where we have introduced the dimensionless quantity
\ba
\tilde\chi&=&\frac{\chi}{\beta\eta_\Phi^2}.
\ea
We also note that the scale for the thickness of the domain wall is given by
\ba
({\rm thickness})^{-1}=a=\frac{1}{2} \, \sqrt{\beta\eta_\Phi^2} \, \sqrt{1-4\tilde\chi},
\ea
and from now on we work in units where $\beta\eta_\Phi^2=1$.\footnote{At this point we note that taking $\beta=\frac{1}{2}$ corresponds to the case taken in \cite{Battye:2008zh}.}
We are now in a position to evaluate some physical properties of the straight kink-vorton, so we start by defining the N\"{o}ther current, $j_\mu$, and charge, $Q$,
\ba
j_\mu&=&-\frac{i}{2}\, \left( \Sigma^\dagger\del_\mu\Sigma-\del_\mu\Sigma^\dagger\Sigma \right),\\
Q/V_z&=&\int\, \mathrm{d}x\, \mathrm{d}y\, j_0=\omega\, \int\mathrm{d}x\, \mathrm{d}y\,|\Sigma|^2,
\ea
where $V_z$ is the volume in the $\underline z$ directions.

The Hamiltonian for (\ref{eq:fieldTheoryFullLagrangian}) is calculated in the standard way, and its spatial integral gives the energy. Taking the analytic solution, and the specified choices of parameters, allows us to derive the energy and charge of the straight domain wall as
\ba
\label{eq:energyPerLength}
E/(L_y \, V_z)&=&{\cal E}/L_y=\frac{8 \, \chi^2-10 \, \chi+5+6 \, (1+4\,\chi)(k^2+\omega^2)}{6\,\beta\sqrt{1-4 \, \chi}},\\
\label{eq:charegPerLength}
Q/(L_y \, V_z)&=&{\cal Q}/L_y=\frac{\omega \, (1+4\, \chi)}{\beta \, \sqrt{1-4 \, \chi}},
\ea
where $L_y$ is the length of the domain wall in the $y$ direction.

Having found the exact form of the straight kinky vortons, we now look to what happens when we bend them into a circle. The idea is that while a condensate-free kink is bound to collapse under its own tension, the presence of a condensate may stabilize the collapse. The belief was that a closed loop with a condensate of definite winding number
\ba
N&=&\frac{2\pi}{\lambda}R=kR,
\ea
without charge, would be enough to prevent collapse \cite{Ostriker:1986xc, Copeland:1987th, Haws:1988ax, Copeland:1987yv}, but a more complete analysis revealed that in fact the angular momentum of the loop, when charge is also present, was the important factor  \cite{Davis:1988jq, Davis:1988ip, Davis:1988ij}.\footnote{The collapse instability for vanishing charge can be seen from \eqref{eq:energyPerLength} by setting $\omega=0$ and using $ \chi = -k^2 = -  N^2/R^2$.}  Even when both charge and winding number are present, one must be careful about making statements of stability, as circular vortons may be unstable against non-circular perturbations, in particular there may be pinch instabilities~\cite{Battye:2008zh}.

\begin{figure}
  	\centering
    	\includegraphics[width=0.5\textwidth]{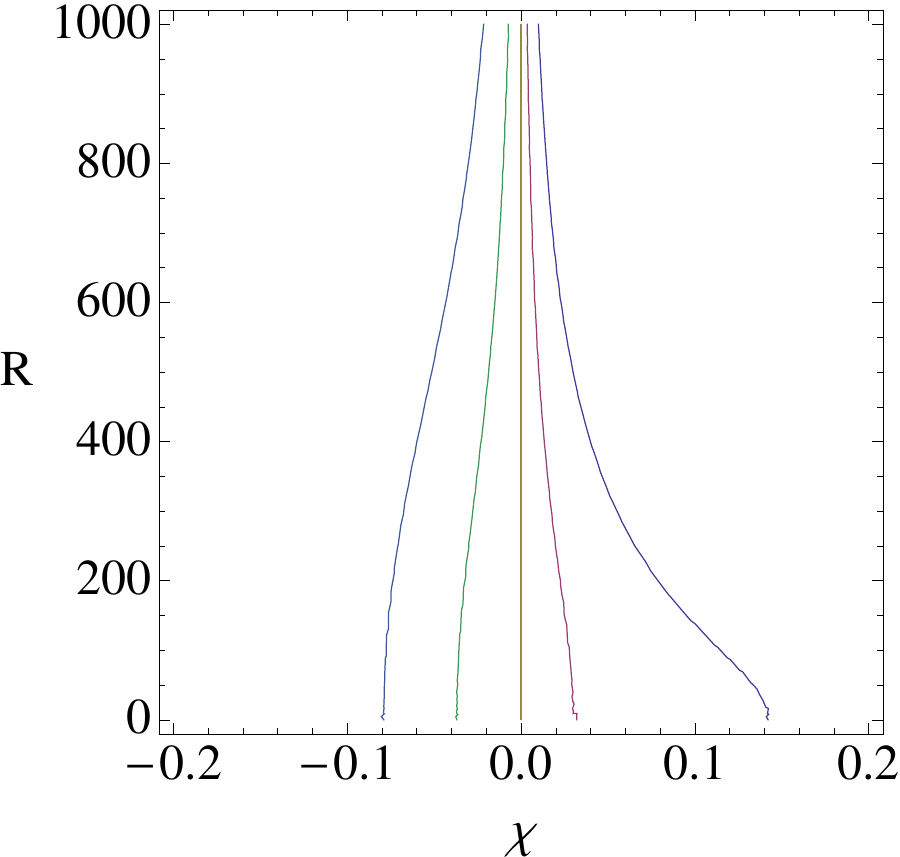}
  	\caption{A plot showing how $R$ varies with $\chi$, for $N=200$ (left), $150$, $\frac{\beta {\cal Q}}{2\pi}$,\;$100$, and $50$ (right) in the case $2\beta{\cal Q}=1500$, in units where $\beta\eta_\Phi^2=1$.}\label{fig:R_chi}
\end{figure}

For this analysis we follow \cite{Battye:2008zh} and consider the approximate solution of a large radius vorton composed with the field profiles of the analytic solution. Taking the square of (\ref{eq:charegPerLength}) for a vorton of radius $R$ ($L_y=2\pi R$), and using $\omega^2=\chi+k^2$ we find that 
\ba
\label{eq:chiEquation}
16\chi^3+8\left( 1+\frac{2N^2}{R^2} \right)\chi^2+\left( 1+\frac{8N^2}{R^2}+\frac{4\beta^2{\cal Q}^2}{4\pi^2R^2} \right)\chi+\left( N^2-\frac{\beta^2{\cal Q}^2}{4\pi^2} \right)\frac{1}{R^2}&=&0 \,.
\ea
In order to get a feel for how $R$ varies with $\chi$ according to this equation we show some examples in Fig.~\ref{fig:R_chi} for fixed winding number and charge.

We will be interested in the thin-wall limit, $R>>1$, which takes us to $\chi\sim0$, as can be seen from Fig.~\ref{fig:R_chi}. In fact, (\ref{eq:chiEquation}) gives a thin-wall approximation to $\chi$ of
\ba
\label{eq:thinWallChi}
\chi_{tw}&\simeq&  \frac{    \left( \frac{\beta {\cal Q}}{2\pi} \right)^2-N^2    }{1+\frac{4\left( \frac{\beta {\cal Q}}{2\pi} \right)^2+8N^2}{R^2}}    \frac{1}{R^2}.
\ea

We now wish to find the energy-minimizing radius, so proceed to calculate the energy by taking (\ref{eq:charegPerLength}) and (\ref{eq:energyPerLength}), which leads to\footnote{Even though we retain the general $\chi$-dependence, it would be consistent to simplify the subsequent discussion by setting $\chi=0$.}
\ba
\label{eq:energyOfVorton}
{\cal E}&=&\frac{2\pi}{\beta\sqrt{1-4\chi}}\left[ \frac{R}{3}\left( 4\chi^2-5\chi+\frac{5}{2} \right)+\frac{1}{R}\left( N^2(1+4\chi)+\left(\frac{\beta {\cal Q}}{2\pi}\right)^2\frac{1-4\chi}{1+4\chi} \right) \right].
\ea
We now minimize this with respect to $R$, taking $\chi$ to be effectively independent of $R$ in the thin-wall limit, to find the minimum-energy vorton radius to be\footnote{Expanding (\ref{eq:energyOfVorton}) to leading order in $\chi$ and using (\ref{eq:thinWallChi}) shows that this is consistent if we take \mbox{$N^2+\left(\frac{\beta {\cal Q}}{2\pi}\right)^2>>N^2-\left(\frac{\beta {\cal Q}}{2\pi}\right)^2$}. In particular, the thin-wall approximation breaks down for vanishing charge, which will have consequences for the applicability of the EFT as will be discussed later.}
\ba
\label{eq:energyMinRadius}
R_0^2&=&3\frac{(1+4\chi) N^2+\frac{1-4\chi}{1+4\chi}\left(\frac{\beta {\cal Q}}{2\pi}\right)^2}{4\chi^2-5\chi+5/2}.
\ea
The final quantity of interest is the angular momentum of the vorton, established in Minkowski spacetime as
\ba
J&=&\int \mathrm{d}\phi\,\mathrm{d}r\, \mathrm{d}z\,r \, T^t_{\;\phi} \nonumber\\
&=&2\pi V_z\int \mathrm{d}r\, r \, 2\,\omega \, k \, |\Sigma|^2 \nonumber \\  
&=&2\, k \, {\cal Q}\,V_z.
\ea
So we introduce the angular momentum per unit $z$-volume ${\cal J}=J/V_z$, to find
\ba \label{eq:angular_mom}
{\cal J}&=&2\frac{N{\cal Q}}{R}.
\ea
We are now in a position to be able to calculate the energy-minimizing radius $R$ and quantity $\chi$ by specifying the winding number $N$ and the charge ${\cal Q}$. This follows by substituting $R_0$ from (\ref{eq:energyMinRadius}) into (\ref{eq:chiEquation}) and solving, numerically, for $\chi$. This $\chi$ is then used in (\ref{eq:energyMinRadius}) to find the radius $R_0$. As this analysis for circular vortons uses the analytic solution for straight kinks, then we only expect this approximation to hold for large radii. For example, \cite{Battye:2008zh} have shown that for winding numbers greater than 10 and radii greater than 50 the approximation and full field theory simulations are in agreement. Importantly, they also found no solutions for a number of cases, such as ${\cal Q}=500$ (or $N=10$). This highlights that the microphysics plays an important role in the existence of these large-scale defects, and so to construct a consistent braneworld model one must take this physics into account. In particular, it was seen in  \cite{Battye:2008zh} that if ${\cal Q}=0$ (or $N=0$), then the vortons have zero radius which, in the context of braneworlds constructed with these domain walls, would rule out braneworlds that are ``stabilized'' by winding alone.
\subsection{Thin wall description}
\label{sec:fieldTheoryVortonsEffective}
We may now take the analytic solution for the vortons and come up with an effective action in the thin-brane limit by integrating the field theory action transverse to the kink,
\ba
S&=&\int{\rm d}x  \int {\rm d}t \,{\rm d}y \, {\cal L},
\ea
where by imposing the ansatz (\ref{eq:analyticProfiles}-\ref{eq:analyticProfiles_sigma}) along with $\Sigma=|\Sigma|\exp\{ i\sigma(t,y) \}$ we may perform the $x$-integration to yield
\begin{subequations}
\ba
S&=&\int{\rm d}t \, {\rm d}y \, \left[  -\frac{1}{2} \, \del_\mu\tilde\sigma\del^\mu\tilde\sigma-\lambda_{(d-1)}\right], \label{eq:effective_theory}\\
\tilde\sigma^2&=&\frac{2 \, (1+4 \, \chi)}{\beta \, \sqrt{1-4 \, \chi}}\sigma^2, \label{eq:renormalized_field}\\
\lambda_{(d-1)}&=&\frac{8 \, \chi^2-10 \, \chi+5}{6 \, \beta\sqrt{1-4 \, \chi}}, \label{eq:tension_chi}
\ea
\end{subequations}
so the kinky-wall looks like a thin wall with a canonical massless scalar, $\tilde\sigma$, living on it.\footnote{Given that $\sigma\sim\sigma+2\pi$ we find that $\tilde\sigma\sim\tilde\sigma+2\pi\sqrt{\frac{2(1+4\chi)}{\beta\sqrt{1-4\chi}}}$.}

The energy of a kink with a condensate is found by calculating the Hamiltonian from the action, and yields
\ba
{\cal E}&=&2\pi R \, \lambda_3+\frac{\pi \, \tilde N^2}{R}+\pi R \, \tilde \omega^2,
\ea
where we have introduced the rescaled $\omega$ and $N$
\ba
\tilde\omega^2&=&\frac{2 \, (1+4 \, \chi)}{\beta \, \sqrt{1-4 \, \chi}} \, \omega^2,\qquad\tilde N^2=\frac{2 \, (1+4 \, \chi)}{\beta \, \sqrt{1-4 \, \chi}} \, N^2,
\ea
and find that the expression for energy matches the previous expression (\ref{eq:energyPerLength}), as it should. In the next section we will use the thin wall description to study the gravitational response within a six dimensional braneworld model.
\section{Kinky braneworlds}
\label{sec:thin_brane}
Here we employ the kinky vorton as a microscopic core model for a
4-brane, $\Sigma_4 = \mathbb{R}^3 \times \mathcal{S}^1$, with three
infinite and one circular spatial dimension. In other words, we
consider the case for which the axial direction corresponds to a three
dimensional manifold [with coordinates $\underline{z}=(z_1,z_2,z_3)$]
describing the spatial dimensions of our universe. Accordingly, the
circular vorton dimension plays the role of a compact brane dimension,
see Fig.~\ref{fig:sketch}.

We are mainly interested in the gravitational response of the coupled
vorton-gravity system. As this is hard to solve analytically, we
employ the effective theory introduced via~\eqref{eq:effective_theory} to describe the system in the
\textit{thin brane} limit. Correspondingly, the worldvolume theory in
its covariant form reads
\begin{align}\label{eq:brane_action}
S_{\mathrm{brane}} = \int \rd^5\tilde{x} \, \sqrt{-\det\left( \tilde{g} \right)} \left[ - \frac{1}{2} \partial_{\tilde{\mu}} \tilde{\sigma} \partial^{\tilde{\mu}} \tilde{\sigma} - \lambda_5 + M_6^4 \Delta K\right]\;,
\end{align}
where the last term is the Gibbons-Hawking-York boundary term, constructed out of the extrinsic curvature $K$.

\subsection{Bulk-brane system}

We start with the exterior, $r>R_0$, bulk metric,
\begin{align}\label{eq:metric}
 \rd s^2_{\mathrm{ext}} = - \, u(r) \, \left[\rd t + a(r) \, \rd \varphi \right]^2 + \re^{2\, k(r) / 3} \rd \underline{z}^2 + F^2 \, \re^{2 \, k(r)} \rd r^2 + r^2 \, u(r)^{-1} \rd \varphi^2 \;,
\end{align}
which adapts the ansatz used by~\cite{Davies} (and first introduced in~\cite{levy_robinson_1964}) to describe a rotating cylinder to 6D.\footnote{To make contact with~\cite{Davies} identify $u = \re^{2\,\bar{u}}$.} The constant
$F$ accounts for the presence of a conical deficit angle. The
corresponding vacuum field equations read
\begin{subequations}
\begin{align}
  2\, F^2 \, \frac{\re^{2 k}}{u} \, R_{tt} &\equiv \frac{1}{ \, r^2}   \left( u \, a' \right)^2 - \left(\frac{u'}{u}\right)^2 + \frac{1}{r} \, \frac{u'}{u} + \frac{u''}{u} = 0 \;,\\
  F^2\,\re^{2 \left( k - u  \right)} \, \left( R_{t\varphi} - a R_{tt} \right) & \equiv a'' - \frac{1}{r} \, a' + 2 \, a' \, \frac{u'}{u} = 0 \;,\\
R_{rr} - 3\, F^2\, \re^{4 k / 3} R_{zz}   & \equiv  \frac{1}{2 \, r^2} \, \left( u \, a' \right)^2 +  \frac{1}{r} \, \left( 2\, k' + \frac{u'}{u} \right) + \frac{2}{3} \left(k'^2 - \frac{3}{4} \, \left(\frac{u'}{u}\right)^2 \right)  =0 \;. \label{eq:k1}
\end{align}
\end{subequations}
The first two equations coincide with the 4D ones, and they are solved
by
\begin{subequations}
  \label{eq:ua}
\begin{align}
  u(r) &=  \frac{ 1 - B \left(|E| \, r\right)^{2 \, A}}{A \sqrt{|B|} \left( |E| \, r\right)^{A-1}} \;, \\
  a(r) & = -\,\frac{A}{E \left[1 - B \, (|E| \, r)^{2A} \right]} + C\;,
\end{align}
\end{subequations}
where $A$, $B$, $C$ and $E$ are integration
constants. Eq.~\eqref{eq:k1} then implies
\begin{equation}
  \label{eq:k2}
4 \, r\, k' \left(3 + r \, k' \right) = 3\,\left( A^2 - 1\right )\;,
\end{equation}
which is different from its 4D counterpart and solved by
\begin{equation}
  \label{eq:k3}
k_{\pm}(r) = \frac{1}{2} \left( - 3 \pm \, \sqrt{3} \sqrt{2 + A^2} \right) \log\left(D \, r\right)\;,
\end{equation}
where $D$ is a constant. Only the branch $ k = k_+ $ is continuously
connected to the trivial solution $k=0$. Later we will see that this branch has
to be chosen in order to describe the geometry of a thin vorton configuration.\footnote{In \cite{Frehland:1971if} it was shown that the above solution can be transformed to Kasner's solution \cite{Kasner}. However, the transformation becomes singular in the static limit (given by $E \to 0$ as we will see later) and hence is not suited to our needs.}

The spacetime region inside the brane uses radial coordinate $\bar{r}$,
and is assumed to be Minkowskian (in accordance with the findings in
\cite{Davies}), which in a non-rotating frame reads\footnote{Note that
  we use the same coordinates $t$, $\varphi$ and $\underline{z}$ in
  the interior and exterior, which can be achieved by a constant re-scaling.}
\begin{align}
\rd s^2_{\mathrm{int}} = - \rd t^2  + \rd \underline{z}^2 + \rd \bar{r}^2 + \bar{r}^2 \rd \varphi^2\;.
\end{align}
Further, by introducing the brane coordinates
$\tilde{x}^{\tilde{\alpha}} = (t, \underline{z},
\varphi)$, we can parametrise the brane induced metric as
\begin{align}\label{eq:induced_metric}
\rd \tilde{s}^2 = -\rd t^2 + \rd\underline{z}^2 + R_0^2 \, \rd\varphi^2 \;.
\end{align}
Continuity of the metric across the brane then requires
\begin{align}
a|_{r=R_0}= 0\,, &&  u|_{r=R_0}= 1\,, && k|_{r=R_0}= 0\,.
\end{align}
which in turn implies
\begin{subequations} \label{eq:cont2}
\begin{align} 
  D &= 1/R_0 \;, \\
B &= \frac{1}{4} \, (R_0 \, |E|)^{-2 \left(A+1 \right)} \left[A - \sqrt{A^2 +4 \, R_0^2 \, E^2 } \right]^2\;, \label{eq:B}\\
C &= \frac{A}{E \left[1 - B \left(R_0 |E| \right)^{2A} \right] }\;. \label{eq:C}
\end{align}
\end{subequations}
The remaining integration constants are fixed in terms of the brane
matter through Israel's junction conditions,
\begin{align}\label{eq:israel0}
  \Delta  K^{\tilde{\mu}}_{\;\tilde{\nu}}\,-\,\delta^{\tilde{\mu}}_{\tilde{\nu}} \, \Delta K \, =& \, \frac{1}{M_6^4}
                                                                                                   \left[ T^{\tilde{\mu}}_{(\tilde{\sigma})\tilde{\nu}} - \lambda_5 \, \delta^{\tilde{\mu}}_{\tilde{\nu}} \right] \;.
\end{align}
Here we may think of the tension $\lambda_5$ combining with the matter
stress-tensor to provide the total brane stress-tensor
$T_{(brane,tot)}^{\tilde{\mu}
  \tilde{\nu}}=T_{(\tilde{\sigma})}^{\tilde{\mu}\tilde{\nu}} \, -
\,\lambda_5 \, \tilde{g}^{\tilde{\mu}\tilde{\nu}}$.  We also
introduced the discontinutiy of the extrinsic curvature across the
brane,
\begin{align}
\Delta  K^{\tilde{\mu}}_{\;\tilde{\nu}} : =  K^{\tilde{\mu}}_{\mathrm{(int)}\tilde{\nu}}\Big|_{\bar{r}=R_0} +   K^{\tilde{\mu}}_{\mathrm{(ext)}\tilde{\nu}}\Big|_{r=R_0}   \;.
\end{align}
We used the convention where the interior normal vector points
outwards and the exterior normal vector points inwards, explicitly
$n_{\mathrm{int}} = \partial_{\bar{r}}$ and
$n_{\mathrm{ext}} = - K \, \re^{k(r)} \partial_r$.  Then the Minkowskian
geometry in the interior implies
$ K^{\varphi}_{\mathrm{(int)} \varphi} = 1/R_0 $ as the only
non-vanishing component. On the other hand, the exterior geometry
(\ref{eq:metric}) has the following non-zero extrinsic curvature
components (evaluated at the brane)
\begin{subequations}
  \label{eq:extrK}
\begin{align}
  K_{(\mathrm{ext})t}^{t} &= - \, \frac{1}{2 \, F}\, u'|_{r=R_0} = \frac{-1 + \sqrt{A^2 + 4 \, R_0^2 \, E^2}}{2 \, F \, R_0}\;,\\
  K_{(\mathrm{ext})\varphi}^{t} &= -\, \frac{1}{2  \, F}\, a'|_{r=R_0} = \frac{R_0 \, E}{F}\;,\\
  K_{(\mathrm{ext})\varphi}^{\varphi} &= -\, \frac{1}{2\, F \, R_0}\, \left[ 2 - R_0\, u' \right]_{r=R_0} =  \frac{-1 - \sqrt{A^2 + 4 \, R_0^2 \, E^2}}{2 \, F \, R_0} \;,\\
   K_{(\mathrm{ext})t}^{\varphi} &= - \frac{E}{F \, R_0}\;,\\
 K_{(\mathrm{ext})1}^{1} &= - \, \frac{1}{3\,F}\,k'|_{r=R_0} = \frac{3 \mp \sqrt{3}\, \sqrt{2 + A^2}}{6\, F\, R_0}\;, 
\end{align}
\end{subequations}
where the upper sign corresponds to the choice $k=k_+$ and the lower
one to $k=k_-$.  The matter field is taken to have the following form
\begin{align}
\tilde{\sigma} = \mu_0^{3/2} \left( \omega \, t + N \varphi \right) \equiv \tilde{\omega}\, t + \tilde{N} \, \varphi \;,
\end{align}
where $N$ is the winding number introduced before and $\mu_0$ a mass
scale, which, according to \eqref{eq:renormalized_field}, is fixed
by the underlying vorton model,
\begin{align} \label{eq:mu0}
\mu_0^3 = \frac{2 \, \left(1 + 4 \, \chi \right)}{\beta \sqrt{1 - 4\, \chi}} \;.
\end{align} 
The brane-induced stress tensor has the following non-zero components,
\begin{subequations}
\begin{align}
T^t_{(\sigma)\;t}&=-\frac{\mu_0^3}{2} \left(  \omega^2 + \frac{N^2}{R_0^2}\right) \;, &
T^t_{(\sigma)\;\varphi}&= \mu_0^3 \, \omega N \;,\\
T^\varphi_{(\sigma)\;\varphi}&=\frac{\mu^3_0}{2} \left(\omega^2+\frac{N^2}{ R_0^2} \right) \;,&
T^\varphi_{(\sigma)\;t}&=- \frac{\mu_0^3}{R_0^2} \, \omega \, N \;,\\
T^z_{(\sigma)\;z}&=\frac{\mu_0^3}{2} \left( \omega^2 -  \frac{N^2}{ R_0^2} \right)\;.
\end{align}
\end{subequations}
It is further convenient to introduce the dimensionless quantities
\begin{align}\label{eq:dimless_params}
 \bar{\lambda} = \frac{2 \, R_0 \, \lambda_5}{M_6^4}\;, && \bar{\omega}^2 = \frac{R_0 \, \mu_0^3 \, \omega^2}{M_6^4} \;, && \bar{q}^2 = \frac{\mu_0^3 \, N^2}{R_0 \, M_6^4}\;,
\end{align}
where $M_6$ is the gravitational scale in the bulk. We are now ready
to evaluate the junction conditions \eqref{eq:israel0}, yielding four
independent equations,
\begin{subequations}
\begin{align}
  \left( t \atop \varphi \right): && \frac{E \, R_0}{F} &= \bar{q}\, \bar{\omega}\;, \label{eq:tt}\\
  6\,\left( z \atop z \right)-2\,\left( t \atop t \right) -2\,\left( \varphi \atop \varphi \right): &&  \frac{4}{F}  &= 4 + 3\, \left(\bar{\omega}^2 - \bar{q}^2\right) - \bar{\lambda}\;,\label{eq:zzmttmvarphivarphi}\\
  \left( t \atop t \right)-\left( \varphi \atop \varphi \right) : &&  \frac{\sqrt{A^2 + 4 \, R_0^2 \, E^2}}{F } &=  \left(1-\bar{q}^2 - \bar{\omega}^2 \right)\;, \label{eq:ttmphiphi}\\
    \left( z \atop z \right): && \frac{2 \, \sqrt{3} \, \sqrt{2 + A^2}}{F} &= \pm \,3 \left( 2 + \bar{\omega}^2 - \bar{q}^2 - \bar{\lambda} \right)\;, \label{eq:zz}
\end{align}
\end{subequations}
where the ``+'' and ``--'' sign in the last equation correspond to
$k=k_+$ and $k=k_-$, respectively.
%
%
We obtain
\begin{subequations}
  \label{eq:KOmegaA}
  \begin{align}
      F &= \frac{E \, R_0}{\bar{q} \, \bar{\omega}}\;, \\
  R_0 \, E &= \frac{4 \, \bar{q} \, \bar{\omega}}{4 - 3\, \bar{q}^2 - \bar{\lambda} + 3\, \bar{\omega}^2} \;,\\
  A^2 &= 16\, \frac{\bar{q}^4 + \left( 1 - \bar{\omega}^2 \right)^2 - 2 \, \bar{q}^2 \left( 1 + \bar{\omega}^2 \right) }{\left[4 - 3\, \bar{q}^2 - \bar{\lambda} + 3\, \bar{\omega}^2\right]^2} \;, \label{eq:A}
\end{align}
\end{subequations}
where $\bar{\omega}$ and $\bar{\lambda}$ have to be chosen such that
$A$ is real.

Since all geometry related integration constants have been fixed, the
remaining equation in~\eqref{eq:zz} provides a constraint on the brane
parameters as specified in~\eqref{eq:dimless_params}. In
particular, it determines the radius $R_0$ once the winding number
$N$, tension $\lambda_5$ and frequency $\omega$ have been fixed. After
using the expressions in \eqref{eq:KOmegaA}, the constraint  (\ref{eq:zz})
reads
\begin{multline}\label{eq:stab}
  \pm \sigma_1 \, \sqrt{17 \, \bar{q}^4 - \left( 8 - \bar{\lambda} \right) \bar{\lambda} - 6 \, \bar{\lambda}\, \bar{\omega}^2 + 17 \, \bar{\omega}^4 - \bar{q}^2 \left( 40 - 6 \bar{\lambda} + 34 \, \bar{\omega}^2 \right) + 8 \left( 3 +\bar{\omega}^2 \right)} \\
  -\sqrt{6}\, \left( 2 + \bar{\omega}^2 - \bar{\lambda} - \bar{q}^2 \right) = 0 \;,
\end{multline}
where $\sigma_1 = \mathrm{sgn}\left( 4 + 3\, \bar{\omega}^2 - 3 \, \bar{q}^2 - \bar{\lambda} \right)$. Again, the ``+'' and ``--'' sign corresponds to $k=k_+$ and $k=k_-$, respectively.

\subsection{Parameter space}\label{sec:param}

As a first sanity check, we try to make contact with the Minkowski
analysis. To that end, we take the decoupling limit ($M_6 \to \infty$)
of the above equation, corresponding to
$\left\{ \bar{q}^2, \, \bar{\omega}^2,\, \bar{\lambda} \right\} \ll
1$. This implies $\sigma_1= 1$ and requires the ``+'' sign in
\eqref{eq:stab}. Later, we will see that this choice leads to a
conical geometry if we depart from the decoupling limit. At first
non-vanishing order, we find
\begin{align}\label{eq:declimit}
\bar{q}^2 + \bar{\omega}^2  - \bar{\lambda} \approx 0 \;,
\end{align}
which after restoring $R_0$, using (\ref{eq:dimless_params}), becomes
\begin{align}
\frac{N^2}{R^2_0} +  \omega^2  -2\,  \mu_0^{-3} \lambda_5 \approx 0 \;.
\end{align}
By using the definitions in~\eqref{eq:charegPerLength},
\eqref{eq:tension_chi} and \eqref{eq:mu0}, we can show that this is
identical to the expression in~\eqref{eq:energyMinRadius},
constituting a nice consistency check of our calculations. Note that the above equation also admits solutions for $\omega =0$ which correspond to $R_0 > 0$. These solutions are not supported by our kinky vorton model which implies a collapse in this case. The mismatch occurs because the thin-wall approximation in \eqref{eq:thinWallChi}, which we used to derive the EFT, breaks down in that particular case. We still include this case in our discussion here, because it allows us to make contact with well-studied conical geometries in the literature.

In the next step, we go away from the decoupling limit and discuss the
solutions of~\eqref{eq:stab} in greater generality.

\subsubsection*{Conical branch}

Here we pick the ``+'' sign in~\eqref{eq:stab}, corresponding to
$k=k_+$ in~\eqref{eq:k3}. This branch is of particular interest as
it admits a consistent decoupling limit. We find two solutions
\begin{align}\label{eq:qsq}
\bar{q}^2_\pm = \frac{1}{11} \, \left[ 8 + 3 \, \bar{\lambda} + 11 \, \bar{\omega}^2 \pm 4 \, \sqrt{4 \left(1 - \bar{\lambda} \right)^2+ 22 \, \bar{\omega}^2} \right]\;,
\end{align}
and for these to be a solution to (\ref{eq:stab}) we find
\begin{subequations}\label{eq:regimeq}
\begin{align}
  3 \, \bar{\lambda} &< 3 + 2 \, \sqrt{6} \, |\bar{\omega}|  \quad \mathrm{for } \quad  \bar{q}^2=\bar{q}^2_- \label{eq:condBrAqm} \;,\\
    3 \, \bar{\lambda} &> 3 - 2 \, \sqrt{6} \, |\bar{\omega}|  \quad \mathrm{for } \quad  \bar{q}^2=\bar{q}^2_+ \;, \label{eq:condBrAqp}
\end{align}
\end{subequations}
where we used  $\sigma_1|_{\bar{q}^2=\bar{q}^2_\pm} = \mp 1$.
\begin{figure}
  \centering
  \subcaptionbox{Conical branch ($k=k_+$) with sub-critical
  ($\bar{q}^2=\bar{q}^2_-$) and super-critical
  ($\bar{q}^2=\bar{q}^2_+$) sub-branch. The dash-dotted line
  approximates the dashed bound in the decoupling limit. \label{fig:param_a}}
{\includegraphics[width=7cm]{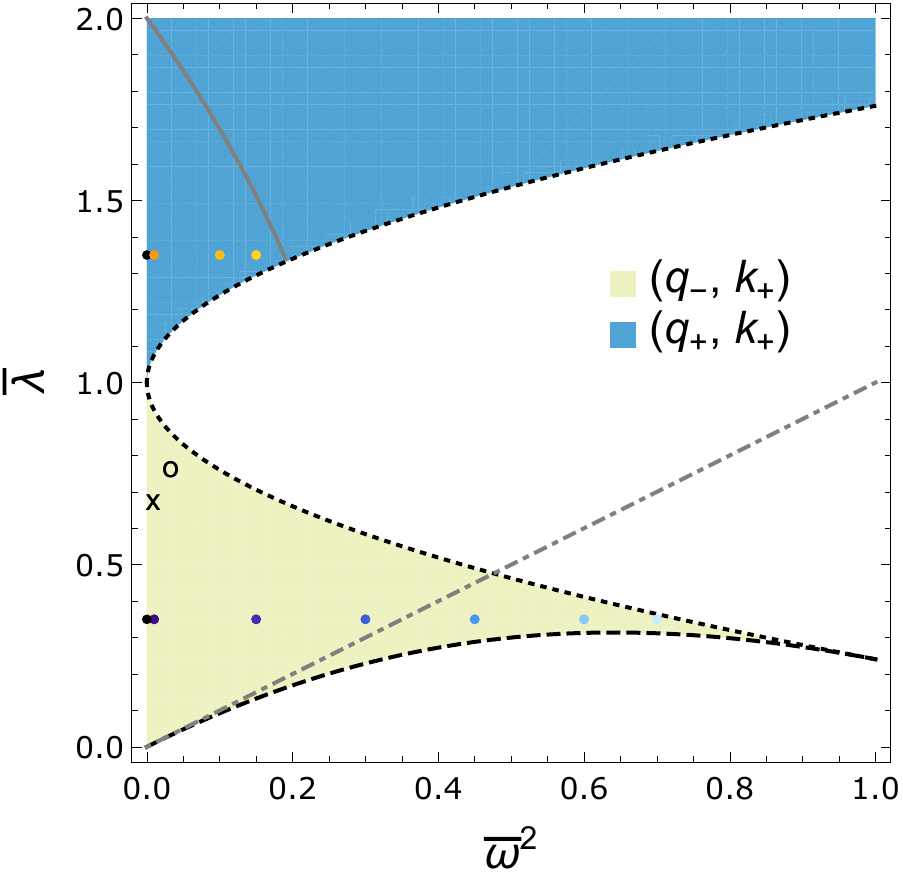}}
\qquad
  \subcaptionbox{Alternative branch ($k=k_-$) with sub-critical
  ($\bar{q}^2=\bar{q}^2_+$) and super-critical
  ($\bar{q}^2=\bar{q}^2_-$) sub-branch. It is incompatible with the
  decoupling limit in~\eqref{eq:declimit}. \label{fig:param_b}}
  {\includegraphics[width=6.7cm]{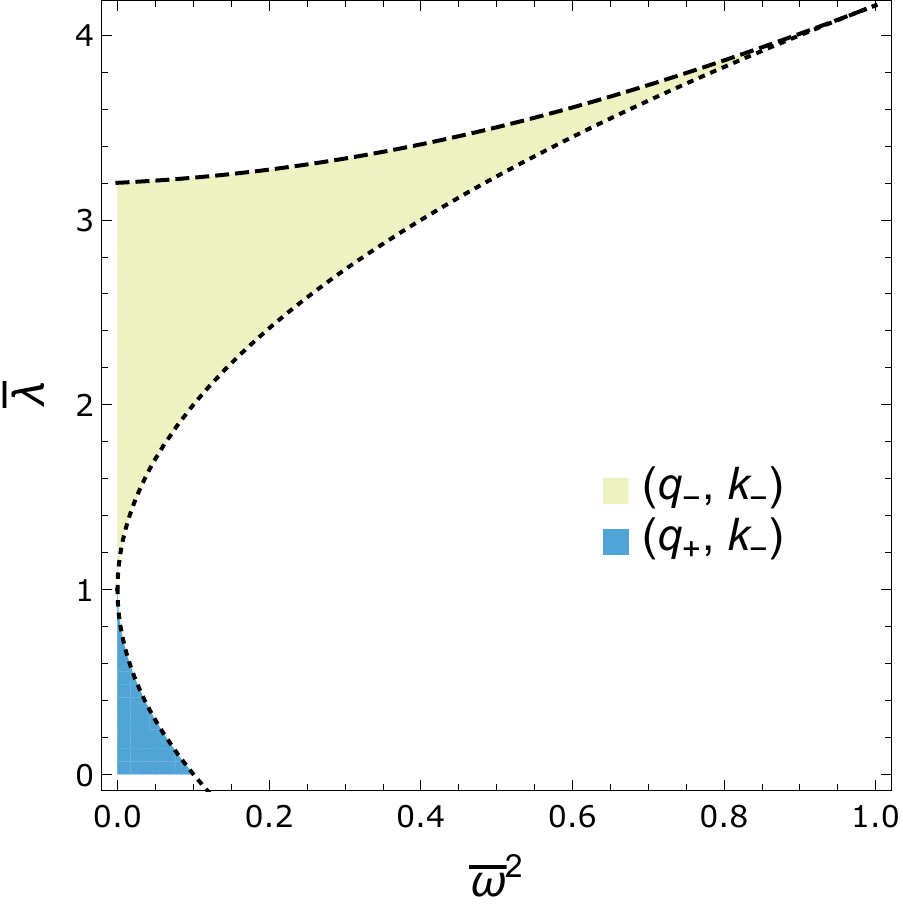}}%
  \caption{Parameter space of the thin brane model. Each pair
    $(\bar \omega, \bar{\lambda})$ corresponds by virtue of
    \eqref{eq:stab} to a particular value of $\bar{q}^2$ (and to a
    certain value of $R_0$ once the winding number has been
    fixed). The dotted and dashed lines depict the positivity bound on
    $A^2$ and $\bar{q}^2$, respectively. Only the shaded regions admit
    a stable solution.}
\label{fig:param_space}
\end{figure}
In order to ensure a stationnary, real solution, we have to make sure
that both $A^2$ and $\bar{q}^2$ are positive, which is not guaranteed
by their respective equations.  For $\bar{q}^2=\bar{q}^2_-$ the positivity of $A^2$ in (\ref{eq:A}) gives an upper bound on $\bar\lambda$, and the positivity of $\bar{q}^2_-$ in (\ref{eq:qsq}) gives a lower bound, leaving us with
\begin{align}\label{eq:BrAbound1}
8 + 3 \, \bar{\omega}^2 - 4 \, \sqrt{2} \, \sqrt{2- \bar{\omega}^2 + 2 \, \bar{\omega}^4} < 5 \, \bar{\lambda} <  5 - 2 \,  \sqrt{3} \, \sqrt{\left(11  - 4 \, \sqrt{6} \right) }\, |\bar{\omega}|\;,
\end{align}
corresponding to the green (light) shaded region in
Fig.~\ref{fig:param_a}. Note that the upper bound in~\eqref{eq:condBrAqm} is weaker.  For ${q}^2=\bar{q}^2_+$ we obtain
a lower bound coming from the positivity of $A^2$ in (\ref{eq:A})
\begin{align}
5 \, \bar{\lambda} >   5 + 2 \,  \sqrt{3} \, \sqrt{\left(11  - 4 \, \sqrt{6} \right) } \,|\bar{\omega}| \;,
\end{align}
which again trumps the one in~\eqref{eq:condBrAqp} and gives rise
to the blue (dark) shaded region in Fig.~\ref{fig:param_a}.

The important message is that the $k=k_+$ branch admits stationnary
solutions with $(\bar{\omega} \neq 0)$ that correspond to a constant
brane radius $R_0$, and therefore consistently generalises the
well-studied static solutions with $\bar{\omega}=0$. The non-shaded
regions, on the other hand, are incompatible with a stabilised radius,
and we therefore expect them to lead to a run-away
behaviour. Moreover, these statements fully take into account the
gravitational back-reaction and are hence applicable to cases of
sizeable 5D energy densities, corresponding to
$\left\{\bar{\lambda},\bar{q}^2, \bar{\omega} \right\} =
\mathcal{O}(1)$, where curvature effects can no longer be
neglected. Specifically, without back-reaction we would have excluded
the parameter regime below the dashed-dotted line in
Fig.~\ref{fig:param_a} to ensure positivity of $\bar{q}^2 $ based
on~\eqref{eq:declimit}, which indeed becomes vastly inaccurate at high
energies.

To obtain a better geometrical understanding of this branch, we
consider the limit $\bar{\omega} \to 0$. We will find that it is
continuously connected to the conical geometry of a static hollow
cylinder with constant surface tension, which is characterised by a
constant deficit angle~$\Delta$. We will henceforth refer to it as the
``conical branch'' (also for $\bar{\omega} \neq
0$). From~\eqref{eq:qsq} and \eqref{eq:regimeq} we find that
$\bar{q}^2 = \bar{\lambda}$ for both $\bar{q}^2=\bar{q}_-^2$ and
$\bar{q}^2=\bar{q}_+^2$ [which also follows from the decoupling limit
in~\eqref{eq:declimit}]. We will refer to them as the ``sub-critical''
and ``super-critical'' sub-branch as they correspond to the disjoint
tension regimes $0<\bar{\lambda}<1$ and $\bar{\lambda}>1$,
respectively. Also note that this result is compatible with the
decoupling limit in \eqref{eq:declimit}, which singles out this branch
as the physically relevant one when we want to describe the geometry
of a kinky vorton in the thin wall limit.

It is straightforward to check that the integration constants in
\eqref{eq:cont2} and \eqref{eq:KOmegaA} reduce to
\begin{align}
  D&=1/R_0 \;, & C &=0 \;, & B &= 1 \;, \nonumber\\
  F &= \left(1 - \bar{\lambda} \right)^{-1} \;, & E&=0 \;, & A&=-1 \;.
\end{align}
Note that~\eqref{eq:A} also admits the solution $A=1$, which we
dismiss as it would imply $C \to \infty$.\footnote{The ``positive A''
  branch might be interesting for non-vanishing values of
  $\bar{\omega}$ though. As we are primarily interested in solutions
  with a continuous limit $\bar{\omega} \to 0$, we will not discuss it
  any further.} We further derive $u(r)=1$ and $a(r)=k(r)=0$, leaving
us indeed with a conical geometry in the exterior,
\begin{equation}\label{eq:geo_static}
  \rd s_{\mathrm{ext}}^2 = - \rd t^2 + \rd \underline{z}^2 \, + \left( 1 - \bar{\lambda} \right)^{-2}\rd r^2 \, + \, r^2 \,\rd \varphi^2 \;,
\end{equation}
where the deficit angle is given by $\Delta = 2 \pi \, \bar{\lambda}$
(see Sec.~\ref{sec:geo} for a more extensive discussion of this
geometry).  Note that for $F>0$ (sub-critical) the range of $r$ is
$[R_0, \infty)$, whereas for $F<0$ (super-critical) it is
$(0,R_0]$.\footnote{This follows from the fact that the normal vector,
  $n= - F \, \re^{k(r)} \, \partial_r$, which is assumed to point in
  the adjacent space, switches its sign when $F$ becomes
  negative. Therefore, to preserve its orientation $r$ has to decrease
  when moving away from the brane.} In the latter case this implies
the existence of a second axis at $r=0$. In summary, the conical
branch corresponds to the choice $k=k_+$, where its sub-critical and
super-critical sub-branch is described by the green (light) and blue (dark) shaded
region in Fig.~\ref{fig:param_a}, respectively.

\subsubsection*{Alternative branch}

Here we briefly discuss the branch with $k=k_-$. The solution for $\bar{q}^2$ is still given by~\eqref{eq:qsq}, only the regimes of validity of the respective
sub-branches have changed,
\begin{subequations}\label{eq:regimeq2}
\begin{align}
  3 \, \bar{\lambda} &< 3 + 2 \, \sqrt{6} \, |\bar{\omega}|  \quad \mathrm{for } \quad  \bar{q}^2=\bar{q}^2_+ \label{eq:condBrBqm} \;,\\
    3 \, \bar{\lambda} &> 3 - 2 \, \sqrt{6} \, |\bar{\omega}|  \quad \mathrm{for } \quad  \bar{q}^2=\bar{q}^2_- \;. \label{eq:condBrBqp}
\end{align}
\end{subequations}
As before we demand positivity of $A^2$ and $\bar{q}^2$, which for
$\bar{q}^2= \bar{q}^2_-$ amounts to
\begin{align}
8 + 3 \, \bar{\omega}^2 + 4 \, \sqrt{2} \, \sqrt{2- \bar{\omega}^2 + 2 \, \bar{\omega}^4} > 5 \, \bar{\lambda} >  5 + 2 \,  \sqrt{3} \, \sqrt{\left(11  + 4 \, \sqrt{6} \right) }\, |\bar{\omega}|\;,
\end{align}
corresponding to the green (light) shaded region in
Fig.~\ref{fig:param_b}. For ${q}^2=\bar{q}^2_+$ we obtain the upper
bound
\begin{align}
5 \, \bar{\lambda} <   5 - 2 \,  \sqrt{3} \, \sqrt{\left(11  + 4 \, \sqrt{6} \right) } \,|\bar{\omega}| \;,
\end{align}
giving rise to the blue (dark) shaded region in Fig.~\ref{fig:param_b}. Like
the conical branch, the solution simplifies considerably in the limit
$\bar{\omega} \to 0$. Specifically, we obtain the following
non-vanishing integration constants: $A=-5$,
$\bar{q}^2 = (16- 5 \bar{\lambda})/11 $ as well as
$F=-11/(1-\bar{\lambda})$, which in turn yields the bulk geometry
\begin{align}
  \rd s^2_{\mathrm{ext}} = \left( \frac{R_0}{r} \right)^4 \, \rd t^2 + \left( \frac{R_0}{r} \right)^4 \, \rd \underline{z}^2 + F^2 \, \left( \frac{R_0}{r}\right)^{12} \, \rd r^2 + r^2 \, \left( \frac{R_0}{r} \right)^{-4}\, \rd \varphi^2 \,.
\end{align}
The existence of this solution had to be expected as a second branch
also exists in the (rotationless) 4D case, typically referred to as
``Melvin'' or ``Kasner''
branch~\cite{Linet1990,Christensen1999}. However, the solution for
$\bar{q}^2$ is obviously incompatible with the decoupling limit in
\eqref{eq:declimit}, so it cannot arise from the microscopic model
considered here and hence will not be considered any
further.\footnote{In fact, for the 4D case it is known that this
  branch imposes a pathological equation of state on the matter
  sector, which strongly questions its physical
  relevance~\cite{Niedermann:2014yka}.}

\subsection{Geometry of rotating braneworlds}
\label{sec:geo}

\begin{figure}
  \centering \subcaptionbox{Radial embedding diagram ($Y=0$): rotation leads
  to a widening of the cone close to the brane. Far away a flat conical profile is approached. \label{fig:embedding}}
  {\includegraphics[width=7cm]{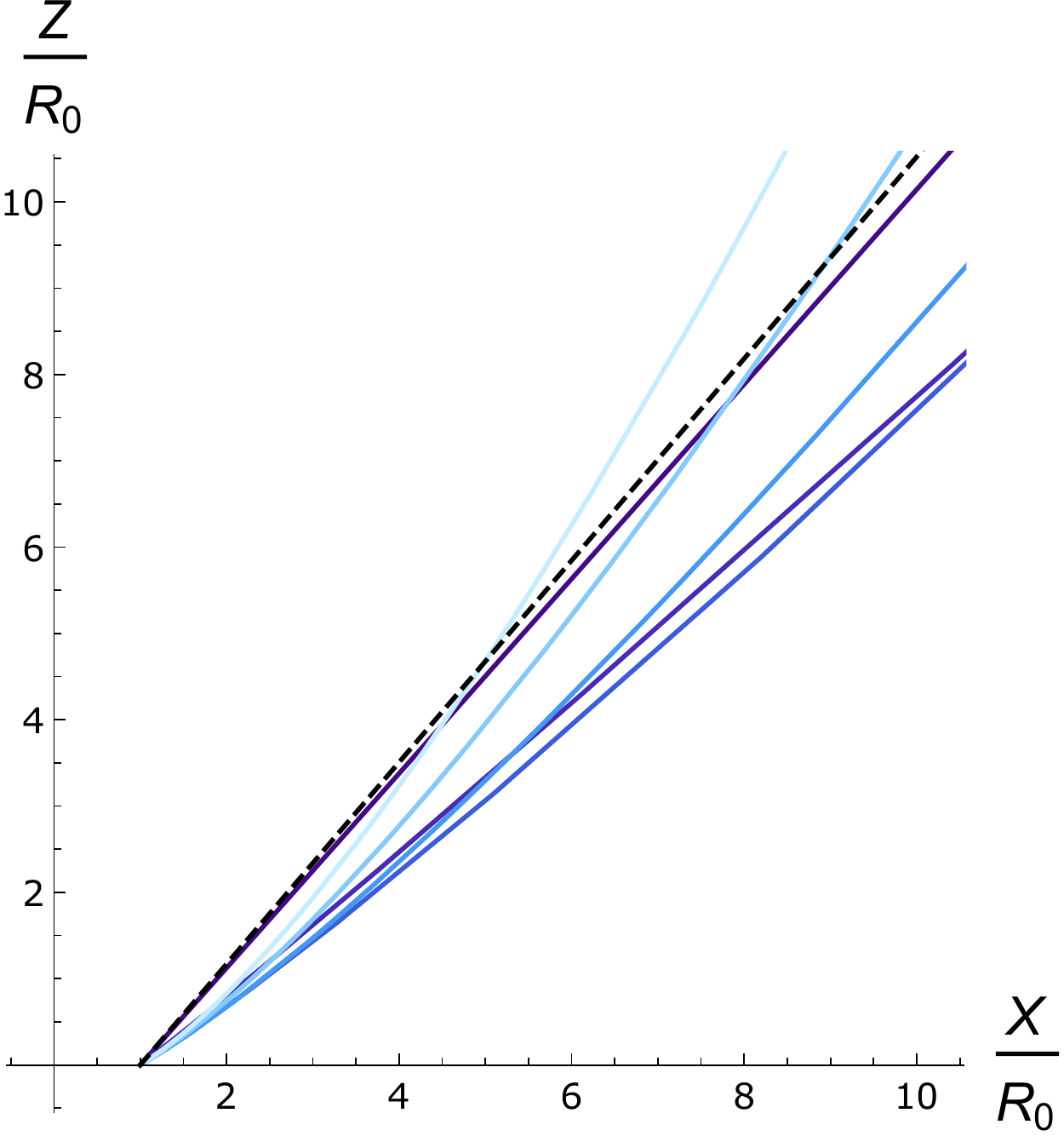}}\qquad
  \subcaptionbox{Deficit angle as ``seen'' by a local bulk observer. Further
  away curvature decreases and a constant deficit angle is approached. \label{fig:deficit}}
  {\includegraphics[width=7cm]{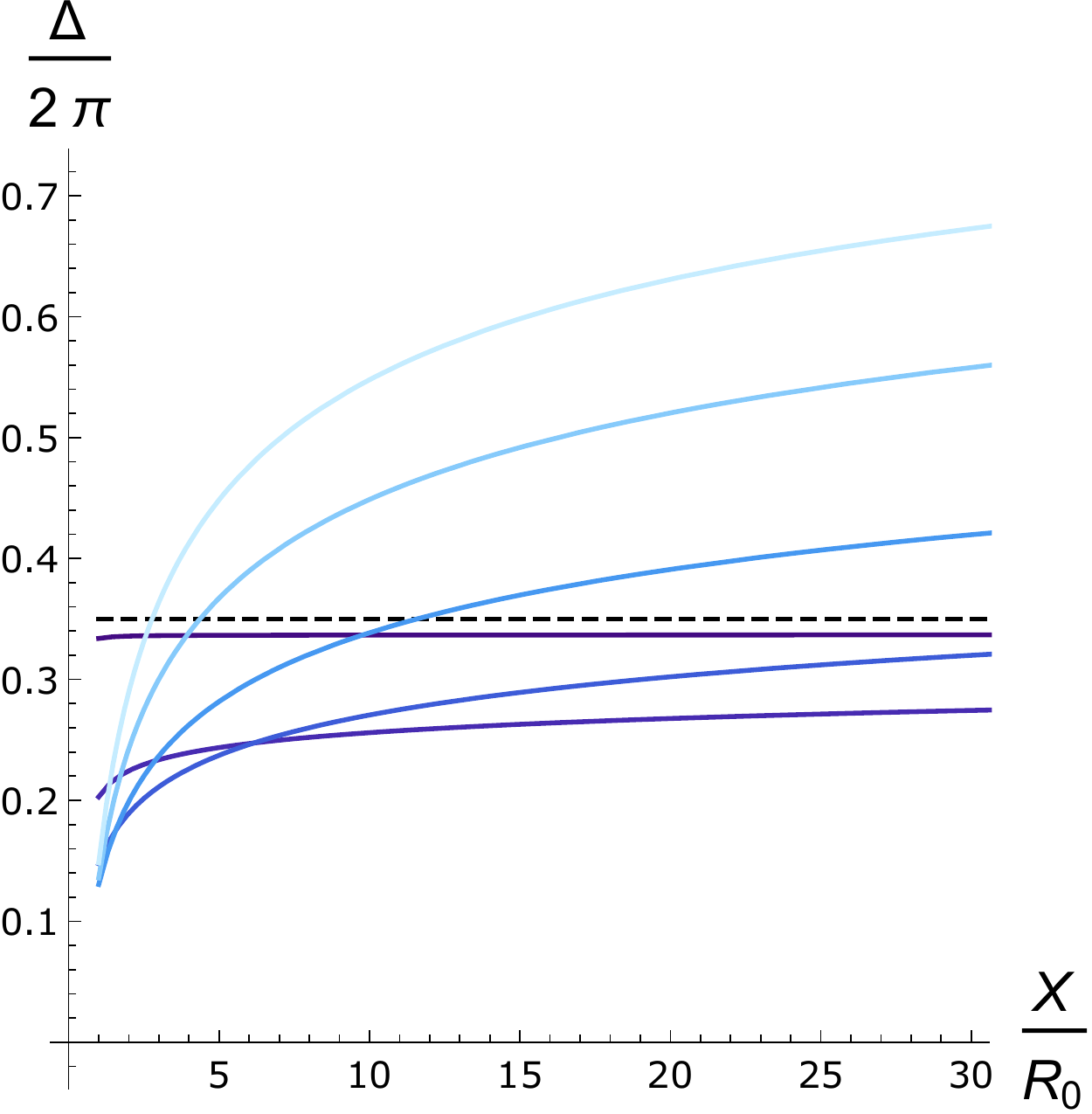}}%
  \caption{Deformed cones for \textit{sub-critical} tension,
    $\bar{\lambda}=0.35$, and
    $\bar{\omega}^2 \in \left\{ 0.01, 0.15,\, 0.3,\, 0.45,\, 0.6,\,
      0.7 \right\}$ [from dark to light, corresponding to the dots in the green (light) region of Fig.\
    \ref{fig:param_a}]. The dashed curve represents the static case
    ($\bar{\omega}=0$). The brane ``sits'' at $X=1$.}
\label{fig:geo_sub}
\end{figure}
In order to infer the geometric impact the parameter $\bar{\omega}$
has on the geometry, it is instructive to construct an embedding
diagram that visualises the extra space curvature. To that end, we
consider a generic extra-dimensional slice of the physical manifold
($\rd t = \rd \underline{z} = 0$),
\begin{align}\label{eq:extra_slice}
\rd s_{\mathrm{ext}}^2 = F^2\, \re^{2 \, k(r)} \, \rd r^2 + v(r)^2 \, \rd \varphi^2  \;,
\end{align}
where we defined
\begin{align}\label{eq:v}
v(r) = \sqrt{\frac{r^2}{u(r)} - a^2(r) \,  u(r) }\;.
\end{align}
This metric can be described as a hypersurface in a three-dimensional
Euclidian space which is parametrised in terms of $(r, \varphi)$
through the embedding functions
$\left[X(r,\varphi),\, Y(r,\varphi),\, Z(r) \right]$, where  $X(r,\varphi)=\cos\left(\varphi\right) \, v(r)$, $Y(r,\varphi)=\sin\left(\varphi\right) \, v(r)$ and $Z(r)$ is determined by the boundary value problem
\begin{align}\label{eq:dgl_embedding}
 Z'(r) = \pm \, \sqrt{ K^2 \, \re^{2 \, k(r)} - \left[v'(r)\right]^2}  \quad \text{and} \quad Z(R_0)=0\;.
\end{align}
For $\bar{\omega}=0$ we can use the static geometry in~\eqref{eq:geo_static}. In that case it is easy to solve the differential
equation, yielding the embedding
\begin{align}\label{eq:embedding_static}
  \left[X(r,\varphi),\, Y(r,\varphi),\, Z(r) \right] = \left[r \, \cos\left( \varphi \right),\, r\,\sin\left( \varphi \right),\, \pm \,\frac{\sqrt{1-\left( 1-\bar{\lambda}\right)^2}}{|1-\bar{\lambda}|} \, \left( r-R_0 \right) \right]\;.
\end{align}
As a further geometrical probe, we introduce the local deficit angle
$\Delta(r)$ a 6D observer would infer from two measurements of the
bulk circumference at $R$ and $R + \rd R$, where $R$ is the proper
radius defined by $\rd R = \sqrt{g_{rr}} \, \rd r $,
\begin{align}\label{eq:deficit_local}
  \frac{\Delta(r)}{2\, \pi} : =  1- \frac{\rd v(R)}{\rd R} 
  = 1- \frac{ v'(r)}{|F| \,\re^{k(r)}}\,. 
\end{align}
The second equality follows from~\eqref{eq:extra_slice}. Evaluated at the brane, we obtain
\begin{align}\label{eq:deficit_local2}
 \frac{\Delta(r)}{2\, \pi} = 1 \mp \left[1 - \frac{1}{8} \, \left( \bar{\omega}^2 + 7\, \bar{q}^2_\mp + \bar{\lambda} \right)  \right]\,,
\end{align}
where we used \eqref{eq:v} and \eqref{eq:extrK}. In the static case ($\bar\omega=0\Rightarrow\bar q^2=\bar\lambda$) this reduces to
$\Delta(r)/(2\pi) = 1 - |1-\bar{\lambda}|$, which agrees with the
well-known expression for a global deficit angle.

\subsubsection*{Sub-critical tension}

\begin{figure}
  \centering \subcaptionbox{Sub-critical tension
  ($\bar{\lambda}=0.35$). Without rotation ($\bar{\omega}^2=0$, left)
  the extra space is a flat cone. With rotation
  ($\bar{\omega}^2=0.45$, right) we observe a slight widening of the
  cone and a build-up of curvature close to the brane.\label{fig:embed_3D_sub}}
  {\includegraphics[width=7cm]{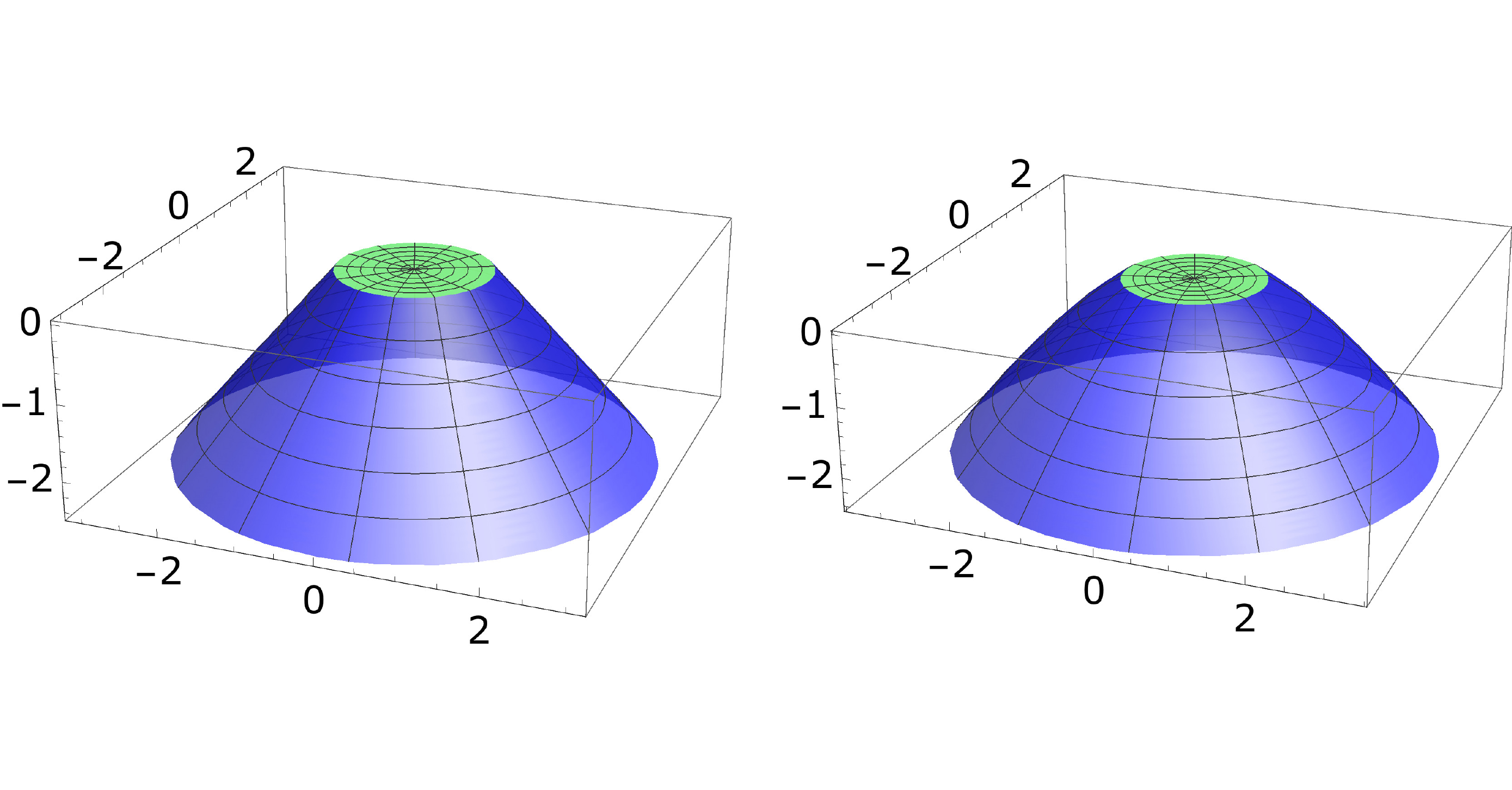}}\qquad
  \subcaptionbox{Super-critical tension ($\bar{\lambda}=1.35$): Without
  rotation ($\bar{\omega}^2=0$, left) the extra space is an inverted
  cone that closes in a conical singularity. With rotation
  ($\bar{\omega}^2=0.001$, right) the extra space is capped at $r=0.65$ (``o'')
  before an excess develops.\label{fig:embed_3D_sup}}
  {\includegraphics[width=7cm]{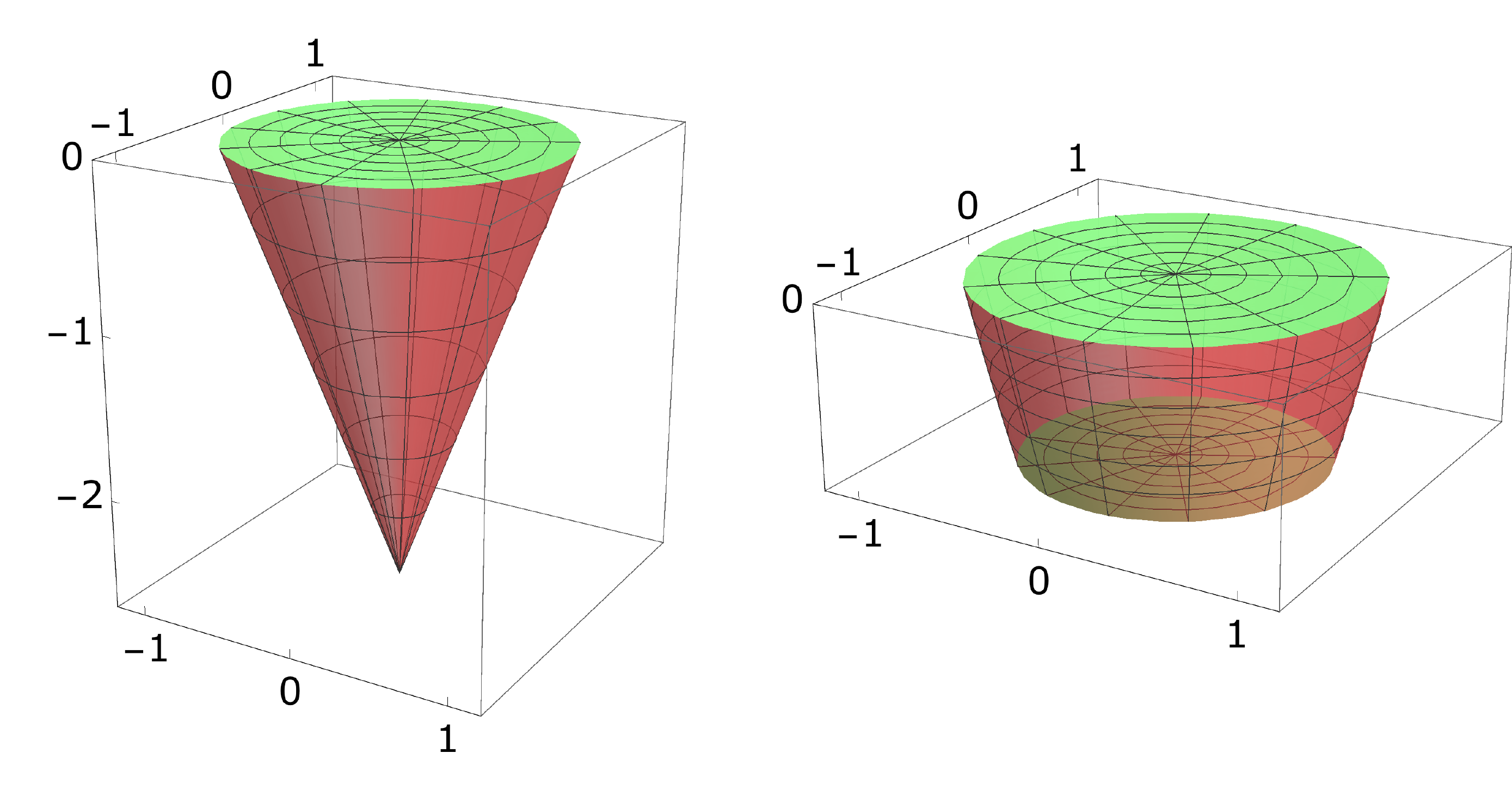}}%
  \caption{Embedding diagrams of the extra space geometry. The
    interior (green) is flat whereas the exterior (blue or
    red) features a conical deficit and is curved for
    $\bar{\omega} \neq 0$.}
\label{fig:embed_3D}
\end{figure}
We first discuss the geometry of a sub-critical tension brane
($\bar{\lambda}<1$). In the static case ($\bar{\omega}=0$) the extra
space is described by a (constant) conical geometry with vanishing
curvature. The corresponding embedding diagramm (for $Y=0$) and
deficit angle are depicted as the dashed line in
Fig.~\ref{fig:embedding} and \ref{fig:deficit}, respectively (see also
Fig.~\ref{fig:embed_3D_sub} for the full angular embedding). This is
the higher dimensional generalisation of the geometry of an infinitely
long, straight cosmic string in
4D~\cite{Vilenkin1981c,Gott1985,Hiscock1985a}. For
$\bar{\omega} \neq 0 $ we integrate~\eqref{eq:dgl_embedding}
numerically and evaluate~\eqref{eq:deficit_local} to obtain the
coloured lines in Fig.~\ref{fig:embedding} and \ref{fig:deficit},
respectively. We see that the angular momentum of the brane leads to a
widening of the cone close to the brane. Contrary to the static case,
the bulk spacetime is no longer flat, it rather has non-vanishing
spatial curvature which becomes strongest in the near-brane region and
falls off as $r \to \infty$. Accordingly, the bulk curves into a
constant cone far way from the brane. Fig.~\ref{fig:deficit} shows
that the asymptotic deficit angle can lie above (for large
$\bar{\omega}$) or below (for small $\bar{\omega}$) the static
value. Further, from the parameter plot in Fig.~\ref{fig:param_a} it
follows that in order to realise a deficit angle that asymptotes to a
near-critical value ($\Delta \lesssim 2 \pi $) we either have to tune
$\bar{\lambda} \lesssim 1$ and $\bar{\omega} \ll 1$ (which is close to
the well-studied static case) or occupy the green (light) sliver around
$\bar{\lambda} \approx 0.2$ which admits $\bar{\omega} \lesssim 1$
(best approximated by the lightest blue line in Fig.~\ref{fig:param_a}).

Another crucial observation is that the intrinsic brane geometry is
flat for all consistent parameter choices, which is obvious from the
induced metric in \eqref{eq:induced_metric}. This implies that the
self-tuning (or degravitation) property, which makes 6D braneworld
models interesting with respect to the cosmological constant problem,
is preserved for our (sub-critical) kinky vorton model (see for example
\cite{Niedermann:2014bqa} and references therein).

\subsubsection*{Super-critical tension}

\begin{figure}%
  \centering
  \subcaptionbox{Radial embedding diagram ($Y=0$): rotation leads to a
  wider inverted cone. To avoid an excess angle, the extra space has
  to be capped by a second (sub-critical) brane sitting to the right
  of the dashed line.\label{fig:embedding_sup}}
  {\includegraphics[width=7cm]{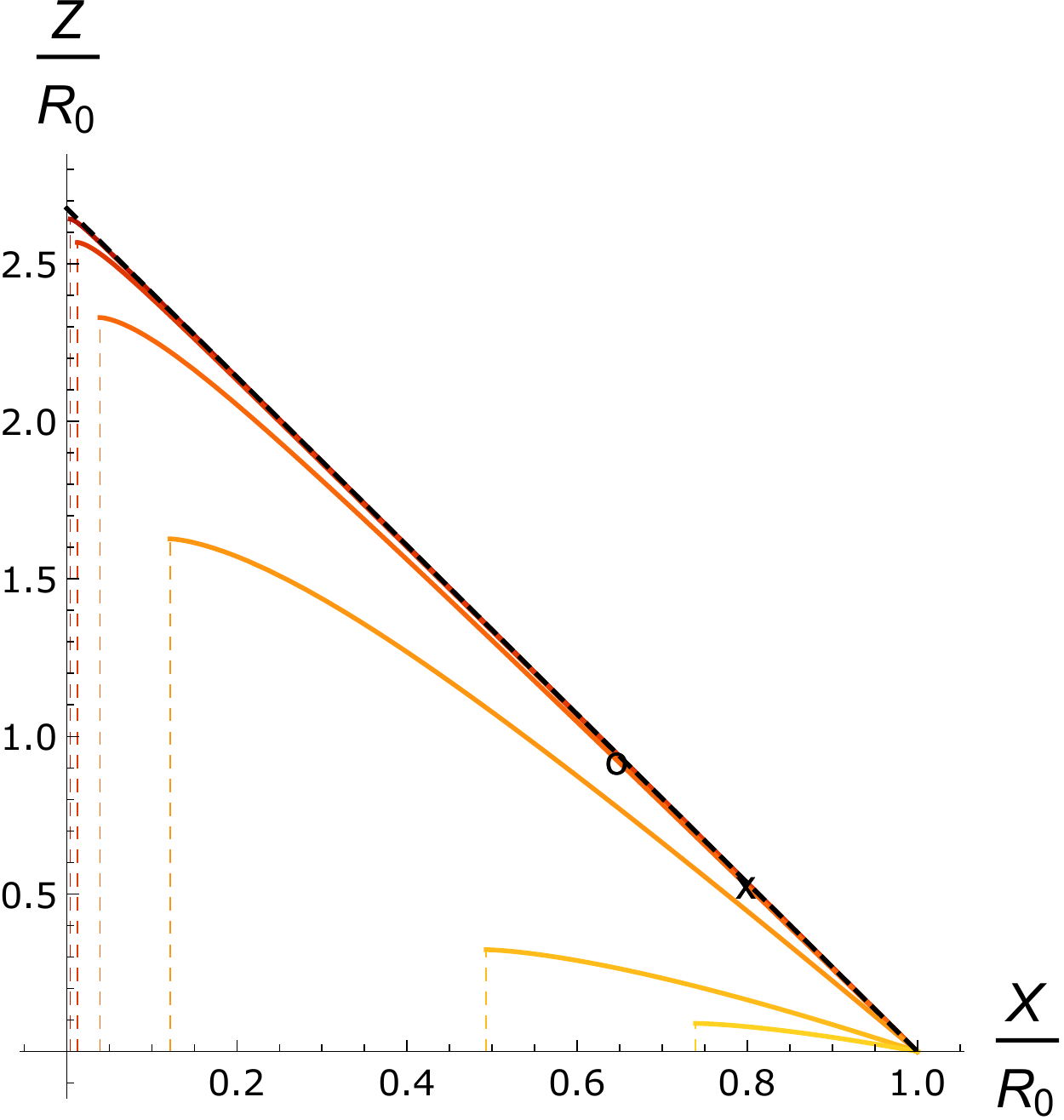}}\qquad
  \subcaptionbox{The local deficit angle decreases away from the brane
  until it turns eventually negative, corresponding to an excess
  angle.\label{fig:deficit_sup}}
  {\includegraphics[width=7cm]{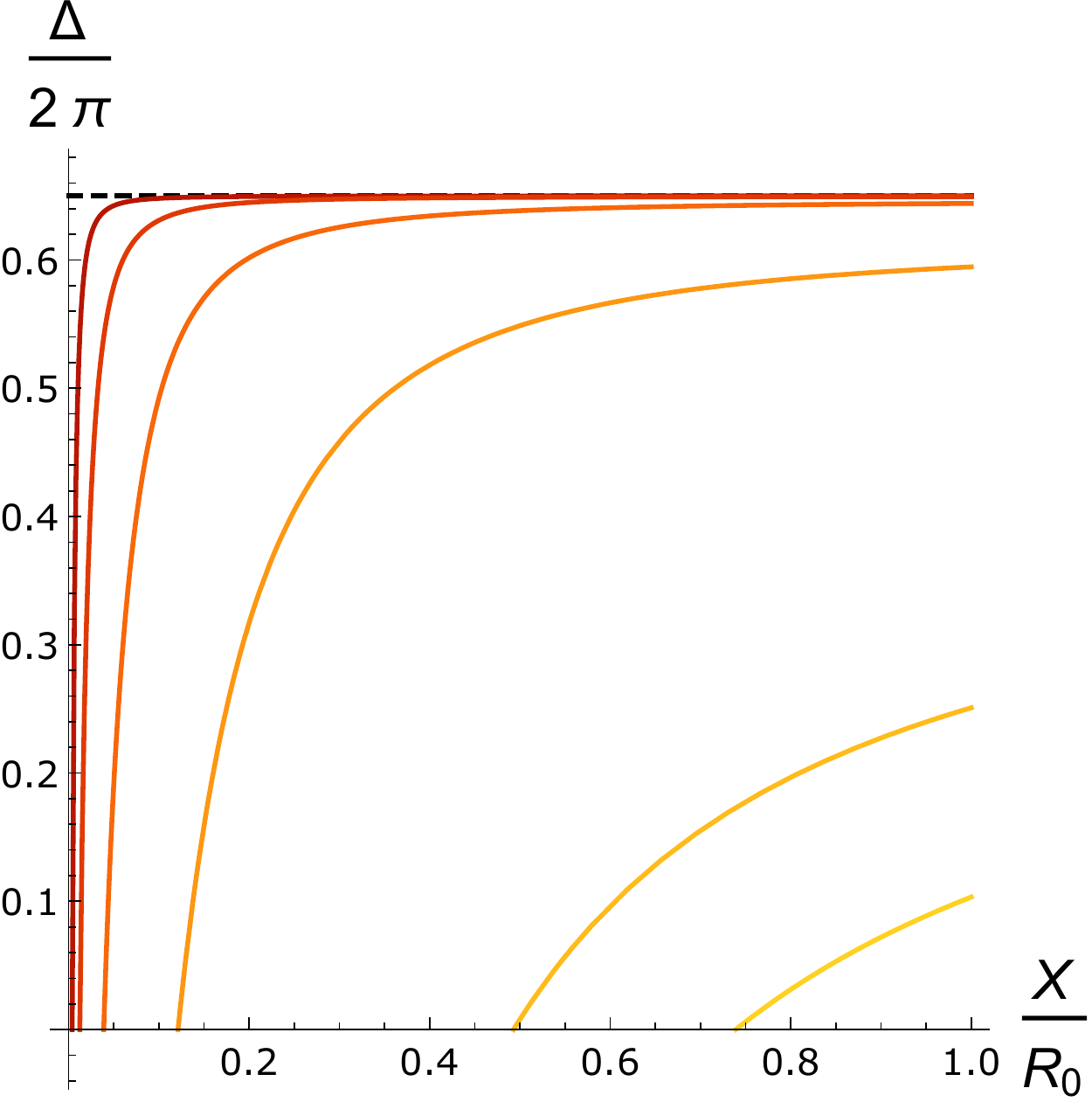}}%
  \caption{Deformed inverted cones for \textit{super-critical}
    tension, $\bar{\lambda}=1.35$, and
    $\bar{\omega}^2 \in \left\{ 10^{-5}, 10^{-4},\, 10^{-3},\,
      10^{-2},\, 10^{-1},\, 0.15 \right\}$ [from dark to light, corresponding to the dots in the blue (dark) region in Fig.~\ref{fig:param_a}]. The dashed curve represents
    the static case ($\bar{\omega}=0$). The brane ``sits'' at $X=1$.}
\label{fig:geo_sup}
\end{figure}
In the super-critical case when $\bar{\lambda}>1$ the circumference of
the extra space shrinks as we move away from the brane. In the static
case, this leads to an inverted cone with deficit angle
$\Delta/ (2 \pi)=2-\bar{\lambda}$, which closes in a second
axis~\cite{Ortiz:1990tn,Blanco-Pillado2014} (at coordinate position
$r=0$), depicted by the dashed lines in Fig.~\ref{fig:geo_sup} (see also
Fig.~\ref{fig:embed_3D_sup} for the full angular embedding). As
this additional axis exhibits a conical singularity, it signals the
presence of second brane with fine-tuned tension
$\bar{\lambda}_* = 2-\bar{\lambda}$. This brane can either be
infinitely thin, giving rise to the observed singularity, or again be
described in terms of an extended configuration, smoothing out the
singularity. In the latter case the
bulk is ``capped'' at some non-vanishing value $0<r_2<R_0$. In both
cases the bulk spacetime becomes compact.

We now move on to the rotating case. We find that the (inverted) cone
is generically widened as we increase $\bar{\omega}$, cf.\ Fig.~\ref{fig:embedding_sup}. Moreover, this effect gets more pronounced
the further we move away from the brane (the smaller $X$ or $r$ is),
and is accompanied by a build-up of spatial curvature. This resonates
with the observation that the curves in Fig.~\ref{fig:deficit_sup}
become steeper as we approach the axis at $X=0$. Eventually, it drops
below zero that way indicating the presence of an excess angle (rather
than a deficit angle). This point is marked by the (vertical) dashed
lines in Fig.~\ref{fig:embedding_sup}.\footnote{Note that there is no
  embedding diagram of an excess geometry, which explains why the plot
  in Fig.~\ref{fig:embedding_sup} cannot be extended beyond the
  dashed line.}  However, as excess angles generically require a
negative brane tension, we dismiss this regime as unphysical (at least
it cannot be described in terms of the kinky vorton model proposed here).
We therefore require the bulk spacetime to be regularised before that
point is reached, which again can be achieved by including a second
brane, cf.\ right plot in Fig.~\ref{fig:embed_3D_sup}. Its
stress-energy has to be tuned such that it gives rise to the correct
value of $\Delta(X)$ at its position. We will provide an explicit
example later in Sec.\ \ref{sec:brane_matching}. We also see from
Fig.~\ref{fig:deficit_sup} that the point where $\Delta(X)$ drops
below zero moves further to the brane for larger values of
$\bar{\omega}$. This implies that there is a maximal $\bar{\omega}$
for which it is no longer possible to regularise the bulk in terms of
a physical, i.e.\ non-negative, brane tension. By demanding
$\Delta(R_0)>0$ we derive from~\eqref{eq:deficit_local} the bound
\begin{align}
 6\, \bar{\lambda} < -2 + 8\, \bar{\omega}^2 + 7\, \sqrt{2} \, \sqrt{2 - 7\, \bar{\omega}^2 + 2\, \bar{\omega}^4}\;,
\end{align}
which is depicted as the grey curve in the blue (dark) region in Fig.~\ref{fig:param_a}. We
thus find that the extra space cone becomes shorter for larger values
of $\bar{\omega}$ due to the necessity of capping the space earlier.

In summary, super-critical solutions ($\bar{\lambda}>1$) can be
consistently generalised to $\bar{\omega} \neq 0$ by introducing a
second sub-critical brane which caps the extra space at a finite
distance away from the axis at $X=0$. We further find that rotation
leads to a widening and shortening of the inverted cone.

\subsection{Dragging of inertial frames}
\begin{figure}%
  \centering \subcaptionbox{ Sub-critical case (static frame): The angular
  momentum of the brane (proportional to $\bar{\omega}$) drags along
  the bulk spacetime with diminishing effect as we move further away. \label{fig:Omega}}
  {\includegraphics[width=7cm]{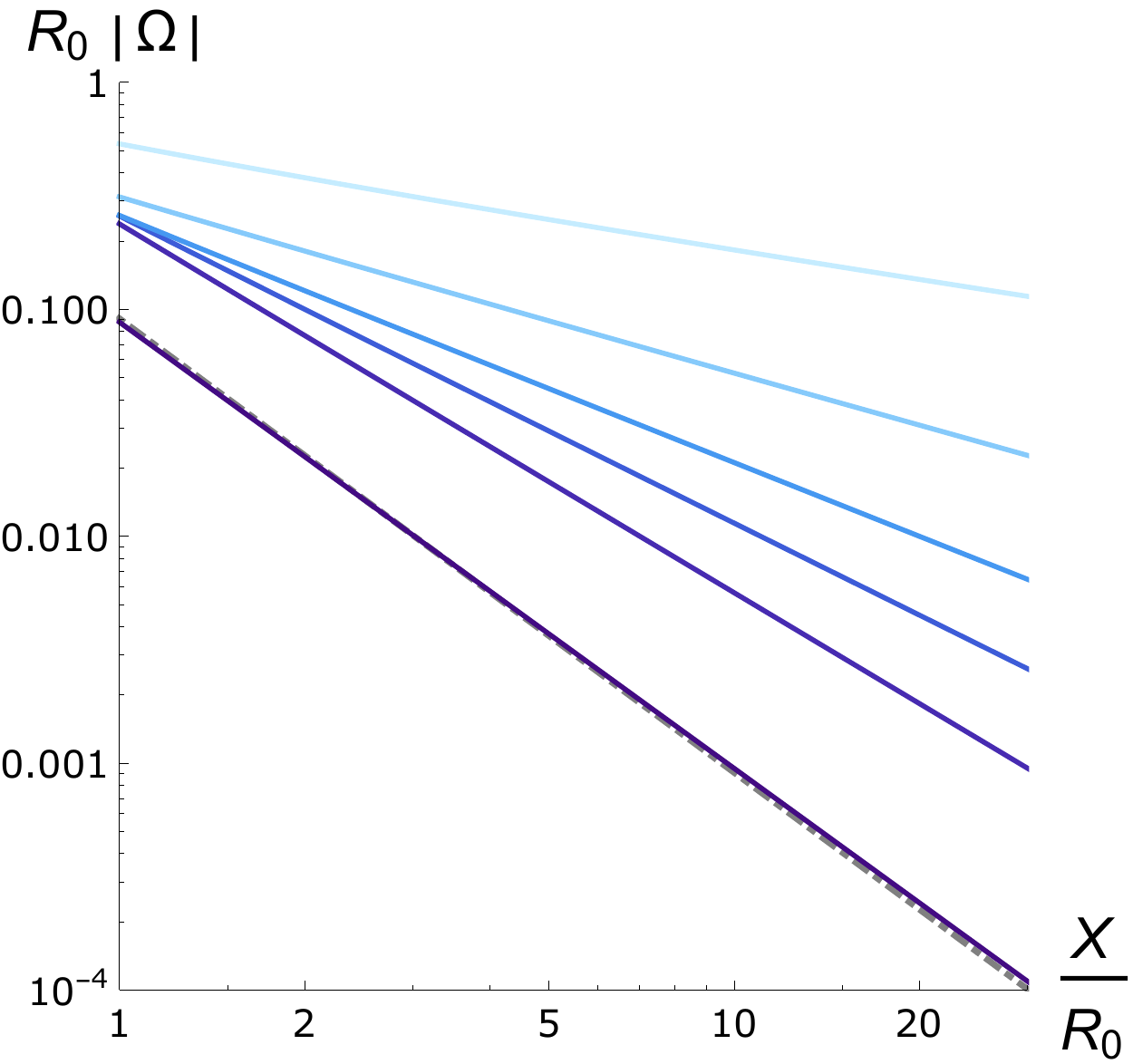}}\qquad
  \subcaptionbox{Super-critical case (co-rotating frame): The ``dragging''
  of the exterior space increases for larger angular momentum of the
  brane.\label{fig:Omega_sup}}
  {\includegraphics[width=7cm]{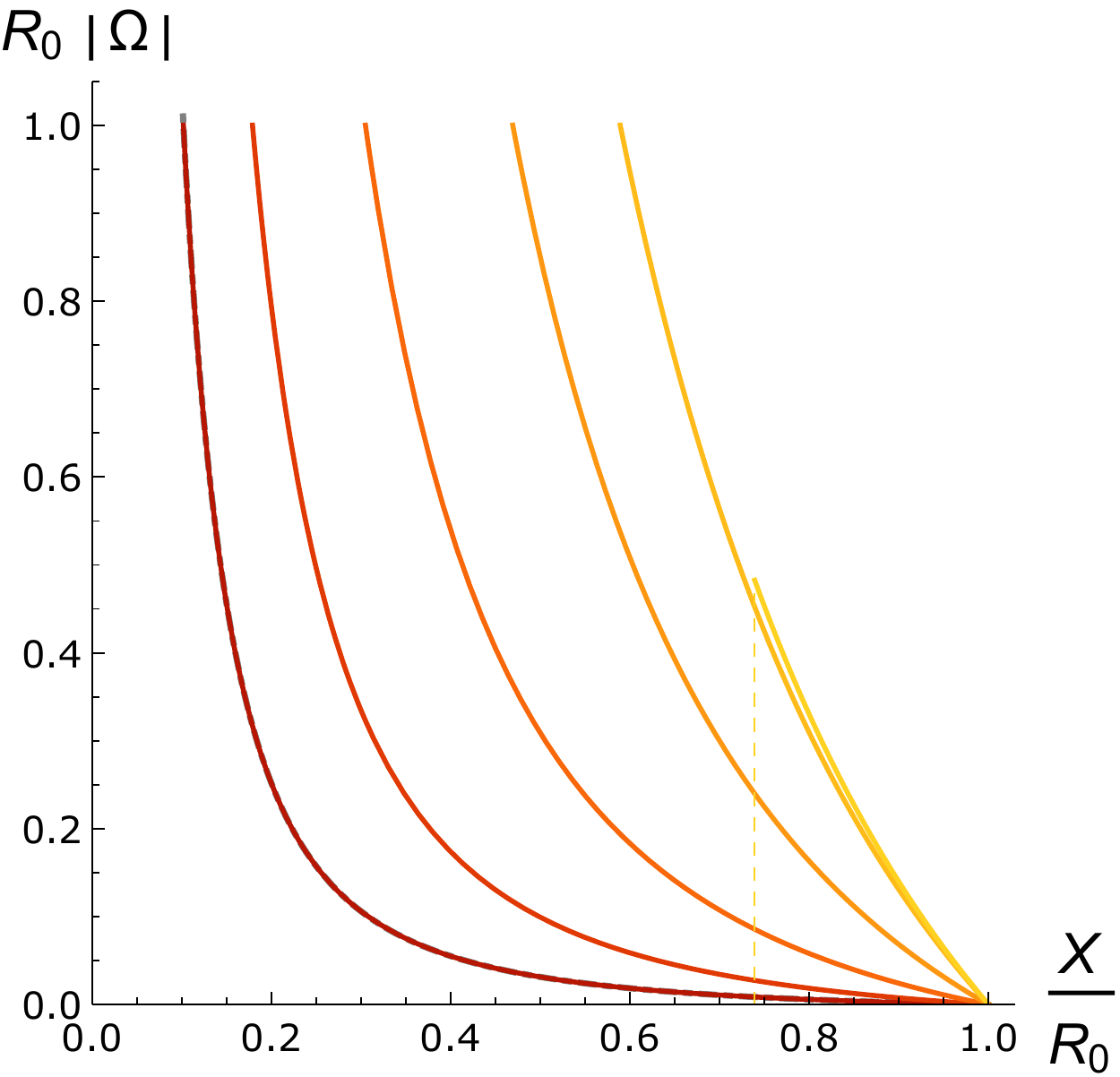}}%
  \caption{Radial profile of the rotation function $\Omega(r)$ with
    parameter values and color coding ($\bar\omega$ increasing from dark to light) as specified in Fig.~\ref{fig:geo_sub} and
    Fig.~\ref{fig:geo_sup}, respectively. The dashed-dotted lines correspond to the analytic result in the slowly rotating case.}
\label{fig:rotation}
\end{figure}
The main new physical feature we introduce in this paper is an angular
momentum of the brane, cf.\ Eq.\ \eqref{eq:angular_mom}. Here we
discuss how this property affects the relative angular motion of
different inertial bulk observers. We first discuss the sub-critical case for
which the bulk is infinite.

We start with the coordinate transformation
\begin{align}\label{eq:angular_trafo}
 \tilde{\varphi} =  \varphi + \Omega_0 \, t, 
\end{align}
where $\Omega_0$ is constant.  The metric \eqref{eq:metric} then reads
\begin{align}
 \rd s^2_{\mathrm{ext}} = -r^{2} \, v(r)^{-2} \, \rd t^2  + \re^{2\, k(r) / 3} \rd \underline{z}^2 + F^2 \, \re^{2 \, k(r)} \rd r^2  + v(r)^2 \left[ \rd \tilde{\varphi} - \Omega(r) \rd t  \right]^2\;,
\end{align}
where $v(r)$ has been defined in \eqref{eq:v}.
We also identified the $r$-dependent function
\begin{align}\label{eq:Omegar}
  \Omega(r) &= \Omega_0 + \frac{a(r) \, u(r)}{v(r)^2} \;.
\end{align}
We further fix the constant $\Omega_0$ by demanding
$\lim_{r \to \infty} \, \Omega(r) =0 $, explicitly
\begin{align}\label{eq:Omega0}
\Omega_0 &= - \, \lim_{r \to \infty}  \frac{a(r) \, u(r)}{v(r)^2} \nonumber\\
 &= E \, \frac{\left[A - \sqrt{A^2 + 4\, R_0^2 \, E^2} \right]^2 - 4 \, R_0^2 \, E^2}{\left(-A \right) \, \left[A - \sqrt{A^2 + 4\, R_0^2 \, E^2} \right]^2}  \;,
\end{align}
where the second line assumes the conical branch.

We will now argue that $\Omega(r)$ provides a sensible measure of the
``rotation of spacetime''. To that end, we consider a slowly rotating
brane, corresponding to
\begin{align}
R_0 \, E \, \approx \, \sqrt{\bar{\lambda}} \, \left( 1 -
  \bar{\lambda} \right)^{-1} \, \bar{\omega} \ll 1 \;.
\end{align}
At linear order in $E$ we find $A \approx -1$, $B \approx 1$,
$C \approx R_0^2 \, E$ as well as $\Omega_0 \approx E$ due to~\eqref{eq:Omega0}. Substituting these into Eqs.~\eqref{eq:ua} and
\eqref{eq:k2} yields $v(r) \approx r$ and $k(r) \approx 0$, which in
turn implies
\begin{align}
 \rd s^2_{\mathrm{ext}} \approx - \, \rd t^2  + \rd \underline{z}^2 + F^2 \rd r^2  + r^2 \left[ \rd \tilde{\varphi} - \Omega(r) \rd t  \right]^2\;,
\end{align}
where due to~\eqref{eq:Omegar}
\begin{align}\label{eq:approxO}
\Omega(r) \approx \frac{R_0^2 }{r^2}\, \Omega_0 \approx \frac{R_0}{r^2} \, \sqrt{\bar{\lambda}} \, \left( 1 -
  \bar{\lambda} \right)^{-1} \, \bar{\omega}\,.
\end{align}
We indeed find that $\Omega(r)$ vanishes in the limit $r \to \infty$,
leading to a non-rotating and locally flat Minkowski metric. The
$(t,\tilde{\varphi})$-frame hence corresponds to a static, inertial
observer at radial infinity (residing at constant $\tilde{\varphi}$)
and thus sets the standard of no rotation. The $(t,\varphi)$-frame, on
the other hand, corresponds to an inertial observer at the brane
(residing at constant $\varphi$). Due to~\eqref{eq:angular_trafo}
[or~\eqref{eq:approxO}], it rotates with respect to the asymptotic
observer with angular velocity $\Omega_0$. The function $\Omega(r)$
generalizes this concept to intermediate observers at radius
$R_0 \leq r < \infty$ and, in that particular sense, corresponds to
the ``rotation of spacetime''.

Eq.~\eqref{eq:approxO} suggests that
the rotation is enhanced when the brane tension approaches the
critical value $\bar{\lambda}=1$. However, from the parameter plot in
Fig.~\ref{fig:param_a} it is clear that this limit is only consistent
if we sent $\bar{\omega} \to 0$ which counteracts the
enhancement.\footnote{Due to the upper bound in~\eqref{eq:BrAbound1}, it is at best possible to achieve a
  constant $\Omega \neq 0$ if the critical limit is taken carefully.}
On the other hand, in the limit where the brane tension is sent to
zero ($\lambda \to 0$), all gravitational effects of the brane
disappear, leading to an empty 6D Minkowski spacetime without rotation.

We now depart from the limit of slow rotation and evaluate the
function $\Omega(r)$ for different values of $\bar{\omega}$ in
Fig.~\ref{fig:Omega}, assuming that it still provides a sensible
measure of rotation. We find a power law
behavior $\Omega(X) \propto X^{- \alpha}$ with $0 < \alpha < 2$, where
the upper limit is approached for a slowly rotating brane in
accordance with~\eqref{eq:approxO} (dash-dotted line).  We also
see that $\Omega$ becomes generically larger, when $\bar{\omega}$ is
increased.\footnote{Close to the brane, non trivial curvature effects
  may affect this simple behavior, which is also visualised in
  Fig.~\ref{fig:geo_sub}.} This nicely resonates with the observation
that the angular momentum of the brane, as defined in the decoupling
limit in~\eqref{eq:angular_mom}, is proportional to
$\bar{\omega}$.

In the super-critical case ($\bar{\lambda}>0$), the extra space is
compact which prevents us from introducing an inertial frame at
radial infinity. We therefore use the super-critical brane as the
non-rotating reference point with respect to which inertial bulk
observers are rotating with angular velocity
\begin{align}\label{eq:Omegar_sup}
  \tilde{\Omega}(r) &= - \frac{a(r) \, u(r)}{v(r)^2} \;.
\end{align}
Note the constant shift in comparison to~\eqref{eq:Omegar}. The
function $\tilde{\Omega}(X)$ is depicted in Fig.~\ref{fig:Omega_sup}
for different values of $\bar{\omega}$. Starting with the
hierarchically small value $\bar{\omega} = 10^{-5}$, we find that
increasing $\bar{\omega}$ leads to an increase in $\tilde{\Omega}(X)$
for all values of $X$, again in accordance with its interpretation as
the rotation of spacetime. All curves diverge towards $X = 0$. Note
however that spacetime has to be cut off before $X=0$ is reached to
avoid an excess angle (for $\bar{\omega}=0.15$ the ``excess point'' is
marked by the dashed line).


In summary, we have argued (and explicitly shown for a slowly rotating
brane) that due to the angular momentum of the brane, inertial
observers at different radial positions are rotating with respect to
each other. This effect, known as the \textit{dragging of inertial
  frames} in the context of rotating black holes, is controlled by the
value of $\bar{\omega}$.

\subsection{Second brane matching}
\label{sec:brane_matching}

We have seen that for a super-critical brane the extra space needs to
be capped by including a second sub-critical brane, otherwise a regime
of (diverging) excess angle close to the symmetry axis occurs. The
crucial question is whether such a brane exists, i.e.\ can be realized
in terms of a physical matter theory. Our analysis allows us to answer
this question, at least if we again employ the thin vorton model to
describe the second brane. To be specific, for every point,
$\left(\bar{\omega}^2,\, \bar{\lambda}\right)$, in the blue (dark) parameter
regime in Fig.~\ref{fig:param_a} (corresponding to the super-critical
brane) and a given radial position $r_* < R_0$, we can ask whether
there exists a dual point,
$\left(\bar{\omega}^2_*,\, \bar{\lambda}_*\right)$, in the green (light)
region (corresponding to a sub-critical brane that consistently caps
the extra space at $r=r_*$).

We first note that for a given branch choice the bulk geometry is
fully determined once the parameters $\bar{\omega}^2$ and
$ \bar{\lambda}$ have been fixed.  It is therefore enough to match two
geometric quantities in order to fix the matter content of the dual
brane. Here we will employ the deficit angle, $\Delta(r)$, and the
rotational profile, $\tilde{\Omega}(r)$. The former is given by~\eqref{eq:deficit_local} and can be (numerically) evaluated at $r=r_*$
for a given choice of $\bar{\omega}^2$ and $\bar{\lambda}$; on the
other hand, it is also related to the matter content on the second
brane according to~\eqref{eq:deficit_local2} subject to the
formal replacement
$\left\{ \bar{\omega}^2, \bar{\lambda} \right\} \to \left\{
  \bar{\omega}^2_*,\, \bar{\lambda}_* \right\}$.\footnote{Note that
  Eq.~\eqref{eq:deficit_local2} is a coordinate independent statement
  and hence holds at \textit{any} brane that is consistently matched
  to the bulk geometry.} Applying the same reasoning to
$R_0^2 \, \tilde{\Omega}'(r)$ as defined in~\eqref{eq:Omegar_sup}, we obtain the two matching equations
\begin{subequations}
  \label{eq:matching}
\begin{align}
  \frac{1}{8} \, \left( \bar{\omega}^2_* + 7\, \bar{q}^2_{*} + \bar{\lambda}_*\right) \, &\stackrel{!}{=} \, \frac{\Delta(r_*)}{2 \pi}\,,\\
  \frac{-8 \, \bar{q}_* \, \bar{\omega}_*}{4 - 3 \, \bar{q}_*^2 - \bar{\lambda}_* + 3\, \bar{\omega^2_*}}\, &\stackrel{!}{=} \, R_0^2 \,\tilde{\Omega}'(r_*)\;,
\end{align}
\end{subequations}
where $\bar{q}^2_* (\bar{\omega}_*, \bar{\lambda}_*)$ is determined by
the `` -- '' branch in~\eqref{eq:qsq}, corresponding to a
sub-critical brane. Note that we used $\tilde{\Omega}'(r)$ [rather
than $\tilde{\Omega}(r)$] as this quantity does not depend on the
choice of the non-rotating reference frame.  As a check of these
relations we can consider the static limit ($\bar{\omega}=0$), which
implies $\bar{\omega}_* = 0 $ and the tuning-relation
$\bar{\lambda}_* = 2 - \bar{\lambda}$ in agreement with the
literature~\cite{Blanco-Pillado2014}. The authors in
Ref.~\cite{Niedermann:2014yka} investigated what happens in cases where the
brane tensions do not fulfil the above relation, and it was found that
stationary solutions exist for which the brane starts to expand in axial
direction with constant rate, corresponding to a 4D de Sitter phase on
the brane. As our ansatz in~\eqref{eq:metric} is not general
enough to accommodate an expanding brane, we remain short on a
definite statement about the rotating brane case. However, it is
conceivable that there is a continuously connected super-critical
solution with $\bar{\omega} \neq 0$ which shows the same inflating
behaviour.

Here we do not provide an exhaustive discussion of the matching and
rather present a proof of existence relying on a special choice of
parameters. Specifically, we use
\begin{align}\label{eq:ex_matching}
  \bar{\lambda} \,= \,1.35\,, && \bar{\omega}^2\, =\, 0.001\,,
\end{align}
corresponding to one of the super-critical solutions depicted in
Fig.~\ref{fig:geo_sup}. We further consider the two radii
$r_*/R_0 \in \left\{0.65,\, 0.8 \right\}$ at which we cap the extra
space, marked by the points ``o'' and ``x'' in Fig.~\ref{fig:embedding_sup} (see also the right plot in
Fig.~\ref{fig:embed_3D_sup}). With this choice we can calculate the
right side of Eqs.\ \eqref{eq:matching}. We then use a numerical root
finding algorithm to determine the corresponding points in the
$\left( \bar{\omega}^2_*,\, \bar{\lambda}_* \right)$-plane, depicted
by ``x'' and ``o'' in Fig.~\ref{fig:param_a}. And indeed we find that
both are within the green (light) shaded region, representing consistent,
stationary (non-inflating) configurations of the super-critical
system. Let us stress that those solutions are particularly
interesting for model building purposes as they correspond to compact
extra dimensions, admitting a 4D gravity regime at low energy scales.

\section{Conclusions}
\label{sec:conclusion}
In this paper we have examined a microphysical model that could lead
(or be extended) to established 6D
braneworlds of
finite~\cite{Peloso:2006cq,Aghababaie:2003wz,Niedermann:2014yka} or
infinite~\cite{Dvali:2000xg,Kaloper:2007ap} extra space volume,
whereby two extra dimensions are hidden by having the Universe live on
a cylindrical brane. In the existing
construction, which used a thin-wall approach to
the calculation, the brane had a massless, periodic scalar field
living inside its worldvolume, and the cylindrical brane was
prevented from collapsing by a winding of the worldvolume scalar. From
the perspective of the microscopic model used here, this
thin-wall approach removes degrees of freedom, which turns out to be
crucial. Indeed, there are unstable modes in the microphysics that
break the rotational symmetry of the extra dimensions, questioning the UV stability of this particular
incarnation of braneworlds. However, using the same model it is
possible to fix the problem. Taking a lesson from the cosmology of
topological defects known as vortons
\cite{Davis:1988jq, Davis:1988ip, Davis:1988ij}, we learn
that rotation of the ring-like object is crucial to prevent its
collapse. In the context of the underlying microphysics, or indeed the
worldvolume theory, this corresponds to having a current circulate
around the loop, rather than simply a winding number.

By using the analytic properties of a field theory model, spelled out
in~\cite{Battye:2008zh, Battye:2009nf}, we have taken parameters from
the microscopic model where the flat-space ring solutions are known to
be stable (under axially symmetric perturbations), and placed them in
a gravitational setting, using a thin-wall approximation. The extra
freedom of rotation gives a richer set of braneworld solutions, where
the rotating braneworld drags the ambient spacetime along with it. We
have explored a range of parameters and have found that in cases where
the deficit angle in the extra dimensions is less than $2\pi$
(sub-critical) we have infinite volume in the extra dimensions, whilst
for larger deficit angles (super-critical) the extra dimensions are
compact.  The former set could be easily extended to the BIG
model~\cite{Kaloper:2007ap} by adding a four dimensional Einstein
Hilbert term to the worldvolume theory in~\eqref{eq:brane_action}. In
fact, this would not change the vacuum configurations as they
correspond to vanishing intrinsic brane curvature. It would then be
interesting whether the new freedom due to rotation helps to avoid the
ghost instabilities diagnosed within the same cylindrical brane setup
in~\cite{Niedermann:2014bqa, Eglseer:2015xla}. Note that, according to
the findings in~\cite{Niedermann:2017cel}, this would require a
screening mechanism (for example Vainshtein
screening~\cite{Vainshtein:1972sx}) to kick in at small distance
scales. On the other hand, the latter set offers, due to its
compactness, a different type of interesting phenomenology.  This
could be further explored by generalizing from a Minkowski worldvolume
geometry to a cosmological setting, just as was done for the cigar
shaped proposal in the non-rotating
case~\cite{Niedermann:2014yka}. Regarding the stability of our
rotating configurations, a final statement still requires a study of
\textit{non-axially} symmetric perturbations. A similar fluctuation
analysis would also allow to infer the phenomenological implications
of the presence of the scalar degree of freedom $\tilde \sigma$, which so far has
only been investigated in the non-rotating case in
\cite{Kaloper:2007ap,Eglseer:2015xla}.

As a final speculation we would like to comment on the effect that
rotation has on fermions bound to the braneworld. In practise, the
braneworld has non-zero thickness, and the braneworld matter fields
have wavefunctions that peak somewhere on the brane.  The precise
location of the wavefunctions of the different fermions has important
consequences for the fermion mass hierarchy, and proton stability, as
pointed out by \cite{ArkaniHamed:1998sj,ArkaniHamed:1999dc}.  This is
due to the overlap between different wavefunctions being exponentially
suppressed as their centres move away from one another. In the context
of our spinning cylindrical braneworld it is natural to expect the
heavier braneworld fermions to be pushed to a larger radius than the
lighter ones, and so could give a natural description of this
mechanism. This would, of course, require a full calculation to give
concrete realisation.

\acknowledgments{FN would like to thank Ruth Gregory
and Paul Sutcliffe for useful discussion. This work was supported by
STFC grant ST/L000393/1.}

\bibliographystyle{JHEP}
\bibliography{library}{}

\appendix

\end{document}